\documentclass[letterpaper]{article} 
\usepackage[]{aaai25}  
\usepackage{times}  
\usepackage{helvet}  
\usepackage{courier}  
\usepackage[hyphens]{url}  
\usepackage{graphicx} 
\usepackage{natbib}  
\usepackage{caption} 
\usepackage{algorithm}
\usepackage{algorithmic}

\usepackage{newfloat}
\usepackage{listings}
\DeclareCaptionStyle{ruled}{labelfont=normalfont,labelsep=colon,strut=off} 
\floatstyle{ruled}
\newfloat{listing}{tb}{lst}{}
\floatname{listing}{Listing}
\usepackage{tabularx}
\usepackage{xspace}
\usepackage{graphicx}
\usepackage{subfigure}
\usepackage{multirow}
\usepackage{fontawesome}
\usepackage{comment}
\usepackage{booktabs}
\usepackage{xcolor}

\newcommand{\NA}{---}

\definecolor{prussianblue}{rgb}{0.0, 0.19, 0.33}

\title{\emph{Why (not) use AI?}\\ Analyzing People's Reasoning and Conditions for AI Acceptability}

\author {
    Jimin Mun\textsuperscript{\rm 1},
    Wei Bin Au Yeong\textsuperscript{\rm 1},
    Wesley Hanwen Deng\textsuperscript{\rm 1},
    Jana Schaich Borg\textsuperscript{\rm 2},
    Maarten Sap\textsuperscript{\rm 1},
}
\affiliations {
    \textsuperscript{\rm 1}Carnegie Mellon University\\
    \textsuperscript{\rm 2}Duke University\\
    jmun@andrew.cmu.edu, wauyeong@andrew.cmu.edu
}

\begin{document}
\maketitle
\begin{abstract}
In recent years, there has been a growing recognition of the need to incorporate lay-people's input into the governance and acceptability assessment of AI usage. 
However, how and why people judge acceptability of different AI use cases remains under-explored, despite it being crucial towards understanding and addressing potential sources of disagreement.
In this work, we investigate the demographic and reasoning factors that influence people's judgments about AI's development via a survey administered to demographically diverse participants (N=197). 
As a way to probe into these decision factors as well as inherent variations of perceptions across use cases, we consider ten distinct labor-replacement (e.g., Lawyer AI) and personal health (e.g., Digital Medical Advice AI) AI use cases. 
We explore the relationships between participants' judgments and their rationales such as reasoning approaches (cost-benefit reasoning vs. rule-based). 
Our empirical findings reveal a number of factors that influence acceptance.
We find lower acceptance of labor-replacement usage over personal health, significant influence of demographics factors such as gender, employment, education, and AI literacy level, and prevalence of rule-based reasoning for unacceptable use cases.
Moreover, we observe unified reasoning type (e.g., cost-benefit reasoning) leading to higher agreement. 
Based on these findings, we discuss the key implications towards understanding and mitigating disagreements on the acceptability of AI use cases to collaboratively build consensus.\footnote{We will release data upon acceptance.} \looseness=-1
\end{abstract}
\section{Introduction}
There is a growing call from the public and experts alike to regulate the development and integration of AI into society \citep{pistilli2023stronger, McClain_2025}. These efforts, as reflected in the EU AI Act \citep{AIAct_2023}, NIST AI Risk Management framework \citep{NIST_2021}, and recent U.S. Executive Order \citep{executiveorder2023}, have resulted in discussions about whether certain AI use cases should be pursued at all. As these high-stakes decisions shape the future of AI, it is essential to equitably determine which AI use cases warrant pursuit despite potential harms, requiring diverse public and expert perspectives. Furthermore, to mitigate potential disagreements, we must understand how individuals make such decisions.

One significant challenge when evaluating the acceptability and impact of AI use cases is that their effects can be simultaneously positive and negative, depending on the context of use, functionality, and broader societal implications \citep{mun2024participaidemocraticsurveyingframework}. For instance, while educational AI can provide affordable and accessible personal tutor, it can, at the same time, lead to over-reliance of students and diminish the goal of education \cite{ChatbotTeach, zhai2024effects}. Thus, as AI is applied across increasingly diverse domains, understanding how people make decisions about its use—especially when benefits and harms conflict—becomes critical for anticipating and addressing disagreements about specific use cases. 

To tackle this challenge, we investigate the factors and reasoning that shape judgments of AI acceptability. First, we assess how judgments about the \emph{acceptability of development and use} vary across different AI use cases,\footnote{By use cases, we mean specific real-world scenarios or problems that an AI system is designed to address.} and how they relate to scenario characteristics (\textbf{RQ1}). Second, we explore \emph{personal factors influencing these judgments}, especially as they relate to demographic differences \cite{kingsley2024investigating} (\textbf{RQ2}). Third, we analyze \emph{reasoning strategies participants use when making judgments} about AI use cases, and how those strategies do or do not relate to the judgments that are ultimately made (\textbf{RQ3}).

\begin{figure*}[hbtp]
    \centering
    \includegraphics[width=\linewidth]{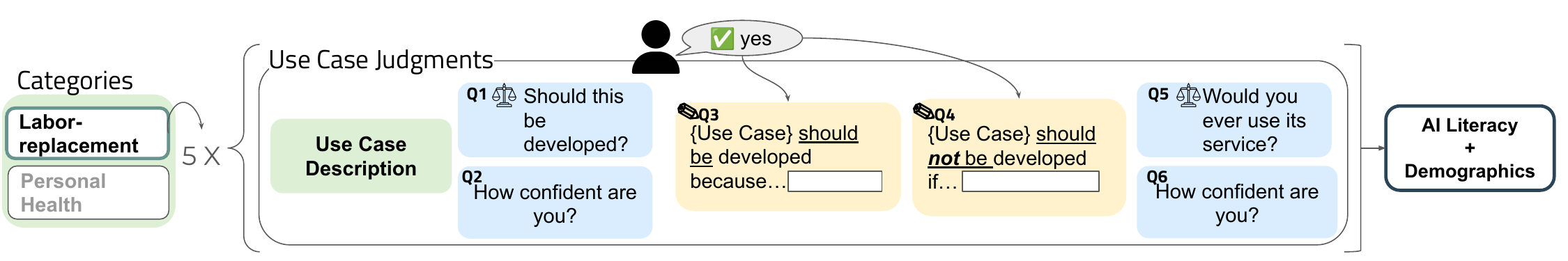}
    \caption{Five professional or personal use cases are presented in a random order. For each use case, we ask multiple-choice questions about its development and confidence levels (Q1, Q2), free-text questions on rationale and decision-switching conditions (Q3, Q4), and multiple-choice questions on usage and confidence (Q5, Q6). These are followed by questions on AI literacy and demographics.}
    \label{fig:survey-flow}
\end{figure*}

To answer these questions, we develop a survey to collect judgments and reasoning processes of 197 demographically diverse participants with varying levels of experience with AI. We ask participants to report whether a certain AI use case should be developed or not, whether they would use such a system, and ask them to provide rationales for their judgment and conditions that would cause them to change their judgments (Figure \ref{fig:survey-flow}). We perform a focused investigation of acceptability using ten AI use cases\footnote{We focus on text-based, non-embodied, digital systems, and while we do not specifically discuss the AI user and subject, in our use case description, we follow three of the five concepts used in EU AI Act to describe high risk use cases \citep{golpayegani2023risk}: the domain, purpose, and capabilities.} that we systematically select for different risk levels, spanning two highly-discussed domains with ongoing efforts to develop such use cases: personal health and labor replacement \citep{McClain_2025,Kelly_2025, Kolata_2024, pierson2025using,rajpurkar2022ai,lee2024contrasting}. To understand characteristics of AI use cases that might affect perceptions beyond category, we vary them by required entry-level education and EU AI risk level (Table~\ref{tab:use-cases}).

We perform a multi-pronged analyses of people's rationales. Drawing from moral philosophy, we examine participants' answers using two reasoning patterns: cost-benefit reasoning, which assesses expected outcomes (e.g., ``using AI for this task would save time''; akin to utilitarian reasoning), and rule-based reasoning, which evaluates the intrinsic values of the action itself (e.g., ``having humans/AI perform this task would be inherently wrong''; akin to deontological reasoning) \citep{cushman2013action,cheung2024measuring}. We then analyze the moral frameworks participants apply, drawing on moral foundations theory \citep{graham2011mapping,graham2008moral}, to identify the dimensions they prioritize in decision making. Finally, to understand conditions under which participants might flip there decisions, we employ three dimensions based on prior studies \citep{solaiman2023evaluating,mun2024participaidemocraticsurveyingframework}: functionality (system capabilities like performance, bias, and privacy), usage (context of system integration, such as supervision, misuse, or unintended use), and societal impact (effects on individuals, communities, and society, such as job loss and over-reliance).

Our empirical results show general higher acceptance of personal health use cases over labor-replacement. While participants' acceptability judgments decreased with increased entry-level education and risk for each category respectively, professional use cases display more variability and disagreements across judgments (RQ1). Acceptability significantly varied among demographic groups and levels of AI literacy, with lower acceptability observed particularly among non-male participants and those familiar with AI ethics (RQ2). Finally, our results show varying distribution of reasoning types across acceptability decisions, with rule-based reasoning being associated with negative acceptance and unified reasoning types showing higher agreement. Further qualitative analysis revealed participants' normative assumptions about AI, humanness, and society—for example, viewing empathy as essential to humanness but lacking in AI (RQ3). \looseness=-1

Our findings shed novel light onto the diversity of people's acceptability and reasoning of AI uses in distinct domains and risk levels. We conclude with a discussion highlighting three key implications for future researchers, practitioners, and policymakers working on advancing ethical and responsible AI development: first, diverse methodologies are needed to effectively analyze use cases and their characteristics; second, involving diverse stakeholders is crucial for assessing the acceptability of AI applications, particularly in workplaces; and third, further investigation into human reasoning processes about AI, notably rule-based reasoning, is needed to inform consensus-building in policy making. \looseness=-1
\section{Related Works}
\label{sec:related-works}

\label{ssec:relatedworks-perception}

While there were many efforts towards ethical AI development and deployment by academics \citep{kieslich2023anticipating, bernstein2021ethics, Neurips2020blog}, industry \citep{openAI_research, deng2024supporting}, and government \citep{NAIRRTF2023FinalReport}, they have largely lacked diverse public inputs.
To address this gap, many works from both academia and civil society have sought to meaningfully engaging lay people in assessing the impact of specific AI use cases. Prior works have focused on anticipating harms \citep{buccinca2023aha} and impacts \citep{kieslich2023anticipating,ada2023survey} through participatory foresight, uncovering diverse, sometimes diverging, viewpoints about AI biases and values \citep{kingsley2024investigating,jakesch2022different,kapania2022because}, and governance efforts of AI \citep{zhang2020us}. However, to the best of our knowledge, only \citeauthor{mun2024participaidemocraticsurveyingframework} considered development decisions by diverse lay-users with an option of not developing a use cases. Among other findings, these prior works from AIES and broader Responsible AI venues revealed a substantial amount of variations in perceptions primarily among demographic lines (e.g., gender, race, political leaning) regarding the desired behavior of AI. 

However, little attention has been given to identifying reasoning of participants over AI use cases. While some works have identified decision variations under ambiguous ethical implications of decisions made by AI for certain tasks (e.g., self driving cars \citep{awad2018moral}, medical AI \citep{chen2023algorithmic}, predictive analysis \citep{barocas2016big}) and inherent value conflicts \citep{jakesch2022different}, these works have not focused on self-reported reasoning. Thus, our work addresses gap by closely examining the \textbf{detailed, self-reported reasoning processes of lay people} regarding the acceptability of AI use cases without explicitly guiding towards outcome-based (i.e., utilitarian) or value-based (i.e., deontological) reasoning, allowing participants to freely choose and express their deliberation process. 

\section{Study Design and Data Collection}
\label{sec:study-design}
To answer our research questions on how and why people judge AI use cases as acceptable, we conducted a survey-based study with demographically diverse participants. In this section, we discuss the selection of use cases (\S~\ref{ssec:use-cases}), survey design (\S~\ref{ssec:survey-design}), and data collection details and participant demographics (\S~\ref{ssec:data-and-demographics}).

\subsection{Use Cases}
\label{ssec:use-cases}
To answer RQ1 which examines the impact of different characteristics AI use cases on judgments and decision making processes, we carefully crafted ten different AI use cases as vignettes. We first chose two broad application categories frequently mentioned by the public in previous works \citep{kieslich2024myfuture,mun2024participaidemocraticsurveyingframework}: \textbf{AI in labor-replacement} where AI takes on a role in society thus far done by a human as a profession (e.g., Lawyer AI), and \textbf{AI in personal health}, where participants could uniformly consider themselves as AI users. We systematically developed five use cases for each category, varying by required education level for labor-replacement applications and by EU AI Act-assigned risk level for personal health applications. 

\begin{table}[!t]
\scriptsize
\centering
\begin{tabularx}{\columnwidth}{l l X}
\hline
Use Case & Factor & Description \\
\hline
\multicolumn{3}{l}{\textbf{Labor-replacement Use}}\\
Lawyer & Doctoral/Prof & Digital legal advice \\
Elementary Teacher & Bachelor & Teaches elementary students \\
IT Support & Some college & Maintains networks; tech support \\
Eligibility Interviewer & High school & Determines benefit eligibility \\
Telemarketer & None & Calls to sell/solicit \\
\hline
\multicolumn{3}{l}{\textbf{Peronal Health Use}}\\
Digital Medical Advice & High Risk & Medical advice pre-consultation \\
Lifestyle Coach & High/Limited & Personalized wellness advice \\
Health Research & Limited & Summarizes personal health info \\
Nutrition Optimizer & Lim./Low & Personalized meal/nutrition tips \\
Flavorful Swaps & Low & Suggests healthy food alternatives \\
\hline
\end{tabularx}
\caption{Study use cases by category. Descriptions are abbreviated. See Appendix~\ref{app:use-case-descriptions} for full descriptions.}
\label{tab:use-cases}
\end{table}
\paragraph{Labor-replacement Use Case Scenarios} 
For the first area of focus, AI in labor replacement, we collected jobs listed in the U.S. census bureau\footnote{https://www.bls.gov/ooh/occupation-finder.htm} and sorted them according to entry level education required as stated in the census. We chose education level as it has been tightly linked to socioeconomic and occupational status \citep{svensson2006professional,evetts2006introduction}. We selected jobs that have a large portion of digital or intellectual components with minimal requirement for embodiment resulting in following five professional roles: Lawyer, Elementary school teacher, IT support specialist, Government support eligibility interviewer, and Telemarketer. See Table~\ref{tab:use-cases} for further details.

\paragraph{Personal Health Use Case Scenarios} 
To understand the acceptability of different health applications in personal and private life, we drew from use cases written by participants from prior works \citep{mun2024participaidemocraticsurveyingframework,kieslich2024myfuture} to systematically craft use cases which varied by risk levels according to EU AI Act. We ensured accurate reflection of the risk levels through iterative refinement of descriptions and agreement with categories assigned by GPT-4, following \citeauthor{herdel2024exploregen}. See Table~\ref{tab:use-cases} for further details.

\subsection{Survey Design}
\label{ssec:survey-design}
Our survey presents participants with five use case descriptions in random order, all from randomly assigned category, labor-replacement or personal health (see \S~\ref{ssec:use-cases} for details). After each description, participants answer: ``Do you think a technology like this should be developed?'' (Q1) and then, ``How confident are you in your above answer?'' (Q2). To allow for examining of their reasoning, participants then provide open-text rationales by finishing the sentence, ``[Use Case] should [not] be developed because...'' (Q3), adjusted dynamically depending on their answer to Q1. 
Participants also described that they would switch their opinion on acceptability of development of the use case: however, ``[Use Case] should [not] be developed if...'' (Q4; also dynamically rephrased based on Q1 answer).
Subsequently, they answer, ``If [Use Case] existed, would you use its service?'' (Q5) and express confidence with, ``How confident are you in your above answer?'' (Q6). Refer to Table~\ref{app:part-1-questions} in the Appendix for the exact wording of the questions.

\paragraph{Collecting Participant Characteristics}
Following the main survey, we asked participants questions about their AI literacy level and demographics to explore various factors affecting perception of AI acceptance (RQ2). We adopted a shortened version of AI literacy questionnaires from previous works \citep{wang2023measuring,mun2024participaidemocraticsurveyingframework} with four AI literacy aspects, \emph{AI awareness, usage, evaluation, and ethics}, and two additional questions for \emph{generative AI, usage frequency and familiarity with limitations}. We collected demographic information of the participants such as \emph{race, gender, age, sexual orientation, religion, employment status, income, and level of education}; see Appendix~\ref{app:demographic-questions} for detailed list of questions. Additionally, we collected information about \emph{discrimination chronicity}, i.e., prolonged experiences of everyday discrimination, of their discrimination experiences (if any) following \citeauthor{kingsley2024investigating}.

\subsection{Data Collection and Participant Demographics}
\label{ssec:data-and-demographics}
We used Prolific\footnote{https://www.prolific.com} to recruit participants. To represent diverse sample, we stratified our recruitment by the ethnicity categories (White, Mixed, Asian, Black, and Other) and age (18-48, 49-100) as provided by Prolific. We also added criteria for quality such as survey approval rating and number of previous surveys completed.
Our study was approved by IRB at our institutions, and we paid 12 USD/hour. Our final sample consisted of 197 participants across two categories, with professional usage assigned to 100 participants and personal to 97. See Appendix~\ref{app:participant-details} for further details on participants.

\section{Acceptability \& Reasoning\\ Analysis Methods}
Our surveys consisted of both multiple choice (numerical) and open-text questions designed to answer our research questions. In this section, we detail our process for numerical (\S~\ref{ssec:num-analysis}) and open-text (\S~\ref{ssec:text-analysis}) analysis.

\subsection{Multiple Choice Analysis}
\label{ssec:num-analysis}
We analyzed the judgment and confidence ratings by mapping judgment (Q1, Q5) to 1 (``Should be developed'', ``Would use'') or -1 (``Should not be developed'', ``Would not use'') and confidence (Q2, Q6) to a scale from 1 to 5. We used numerically converted judgment, confidence, and combined (judgment$\times$confidence; -5 to 5) values as dependent variables in our analysis. We used repeated-measures ANOVAs to understand the differences in mean responses between conditions/groups and linear mixed effects regression models (lmer) to better understand the effects of specific factors. We included a subject-specific random effect when using ANOVA and regression models and added a use-case-specific random effect when applicable. We factorized demographic responses for analysis with the exception of discrimination chronicity, which we aggregated to a numerical value \citep{kingsley2024investigating, michaels2019coding}. We also converted responses to AI literacy questions to numerical values for analysis. 
\subsection{Open-response Analysis}
\label{ssec:text-analysis}
\paragraph{Background: Moral Decision Making}
\label{ssec:moral-decision-making}
To understand decision-making in AI use cases, we draw on moral psychology and dual system theory. We examine two decision-making systems: cost-benefit reasoning, which assesses outcomes and consequences, and rule-based reasoning, focusing on norms, rules, and virtues \citep{cushman2013action,cheung2024measuring}. These correspond to utilitarian reasoning (maximizing good) and deontological reasoning (adhering to moral duties and rights), respectively. Additionally, we apply moral foundations theory \citep{graham2008moral} to identify values and potential moral conflicts in AI development. \looseness=-1

To assess the reasoning methods used by the participants, we analyzed the open-text responses on elaborations to their decisions (Q3) and circumstances in which their decisions would switch (Q4) along the following three dimensions: reasoning types (cost-benefit, rule-based, both, unclear), reference to moral foundations\footnote{We used the five foundational dimensions: Care, Fairness, Loyalty, Authority, and Purity. Although these dimensions have been updated to encompass a broader range of values beyond WEIRD (White, Educated, Industrialized, Rich, and Democratic) populations \citep{atari2023morality}, we selected this version for survey brevity.} (Care, Fairness, Purity, Authority, Loyalty), and switching conditions (Functionality, Usage, Societal Impact). By analyzing reasoning types and moral values reflected in the participants' justifications, we aim to characterize \emph{how} participants made their decisions, and by analyzing various factors such as primary concerns in switching condition, we aim to discover \emph{what} aspects were salient for the participants in their decisions. \looseness=-1

\paragraph{Classification and Aggregation} 
We classified participants' responses to Q3 (elaboration of judgment) and Q4 (conditions for switching decisions), totaling 985 samples for each question, using OpenAI's gpt-4o\footnote{\texttt{gpt-4o-2024-11-20}}. To validate the model's classification performance, results were compared with a reference set of 100 samples annotated by three independent annotators, comprised by members of the research team and a professional annotator. Initially, each annotator independently assessed the data, and then consensus was reached through discussion to establish a gold standard set. The inter-rater agreement between the gold standard and o1-mini's annotations was evaluated using Gwet's AC1 metric, chosen for its robustness with infrequent labels \citep{wongpakaran2013comparison}. While the agreement levels varied, ranging from almost perfect to moderate (0.98--0.57), all dimensions had above substantial agreement except Societal Impact. Annotations for cost-benefit reasoning, rule-based reasoning, and authority reached near-perfect agreement.
Due to minimal occurrences in both human and LLM annotations, the moral foundation dimension Loyalty was excluded from further analysis. The annotations were conducted based on presence or absence of the values and were converted into binary format for statistical analysis. See Appendix~\ref{app:open-text-annotation-details} for further details on agreement and automatic annotation settings.\looseness=-1
\begin{figure*}[!h]
    \centering
    \includegraphics[width=\linewidth]{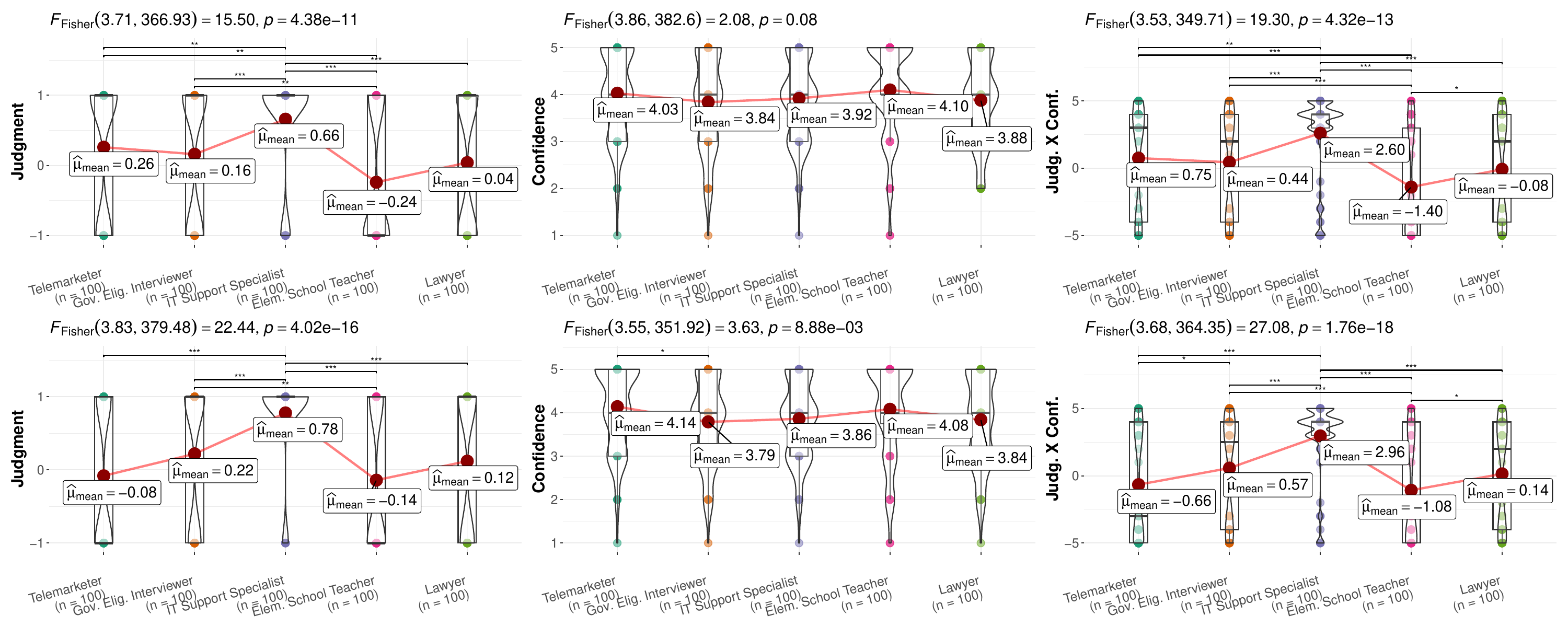}
    \caption{Labor-replacement use case means and distributions of numerically converted Judgment, Confidence, and Judgment$\times$Confidence. First row shows results for development decisions (Q1, Q2) and second row shows results for usage decisions (Q5, Q6). ANOVA results for use cases are shown above each panel. Within subject test was performed using Student's t-test with Holm correction. * denotes following significant p-values: $^{***}p<0.001$; $^{**}p<0.01$; $^{*}p<0.05$. Use case names were shortened.}
    \label{fig:use-case-effect-jobs}
\end{figure*}
\section{Findings}
Our work aims to uncover variations in acceptability of AI use cases and factors and reasoning processes that underlie these judgments. In this section, we discuss our findings about the judgments of the AI use cases (\S\ref{sec:RQ1}), personal factors that may influence the decision such as demographics and AI literacy (\S\ref{sec:RQ2}), and factors in rationales that could uncover reasoning processes that lead to judgments (\S\ref{sec:RQ3}).

\subsection{RQ1. Use Case Perceptions \& Disagreements}
\label{sec:RQ1}


In our analysis, we investigated the effects of use cases on participants' judgments using our ten use case vignettes. Overall, acceptability statistically differed among the two categories ($t_{\texttt{DEV}}(983) = -9.05, p<.001$; $t_{\texttt{USAGE}}(983) = -5.50, p<.001$). Notably, personal health use cases had higher acceptability ($M_{\texttt{DEV}}=0.68, SD_{\texttt{DEV}}=0.74$; $M_{\texttt{USAGE}}=0.51, SD_{\texttt{USAGE}}=0.86$) than labor-replacement use cases ($M_{\texttt{DEV}}=0.18, SD_{\texttt{DEV}}=0.99$; $M_{\texttt{USAGE}}=0.18, SD_{\texttt{USAGE}}=0.98$). See Figure~\ref{fig:use-case-effect-category} in Appendix~\ref{app:extended_results} for additional category comparison results. 

\paragraph{Labor-replacement Use Cases}
Exploring specific use cases within the labor-replacement category (Figure~\ref{fig:use-case-effect-jobs}), we observed that Elementary School Teacher AI ($M_{\texttt{DEV}}=-0.24, SD_{\texttt{DEV}}=0.98$; $M_{\texttt{USAGE}}=-0.14, SD_{\texttt{USAGE}}=1.00$) had the lowest acceptability for both types of judgments followed by Lawyer AI ($M_{\texttt{DEV}}=0.04, SD_{\texttt{DEV}}=1.00$) and Telemarketer AI ($M_{\texttt{USAGE}}=-0.08, SD_{\texttt{USAGE}}=1.00$) where their near zero mean suggest disagreement within judgments. Interestingly, IT Support Specialist AI had the highest acceptability ($M_{\texttt{DEV}}=0.66, SD_{\texttt{DEV}}=0.76$; $M_{\texttt{USAGE}}=0.78, SD_{\texttt{USAGE}}=0.63$) despite human replacement potential and median education level. 

\begin{figure*}[!h]
    \centering
    \includegraphics[width=\linewidth]{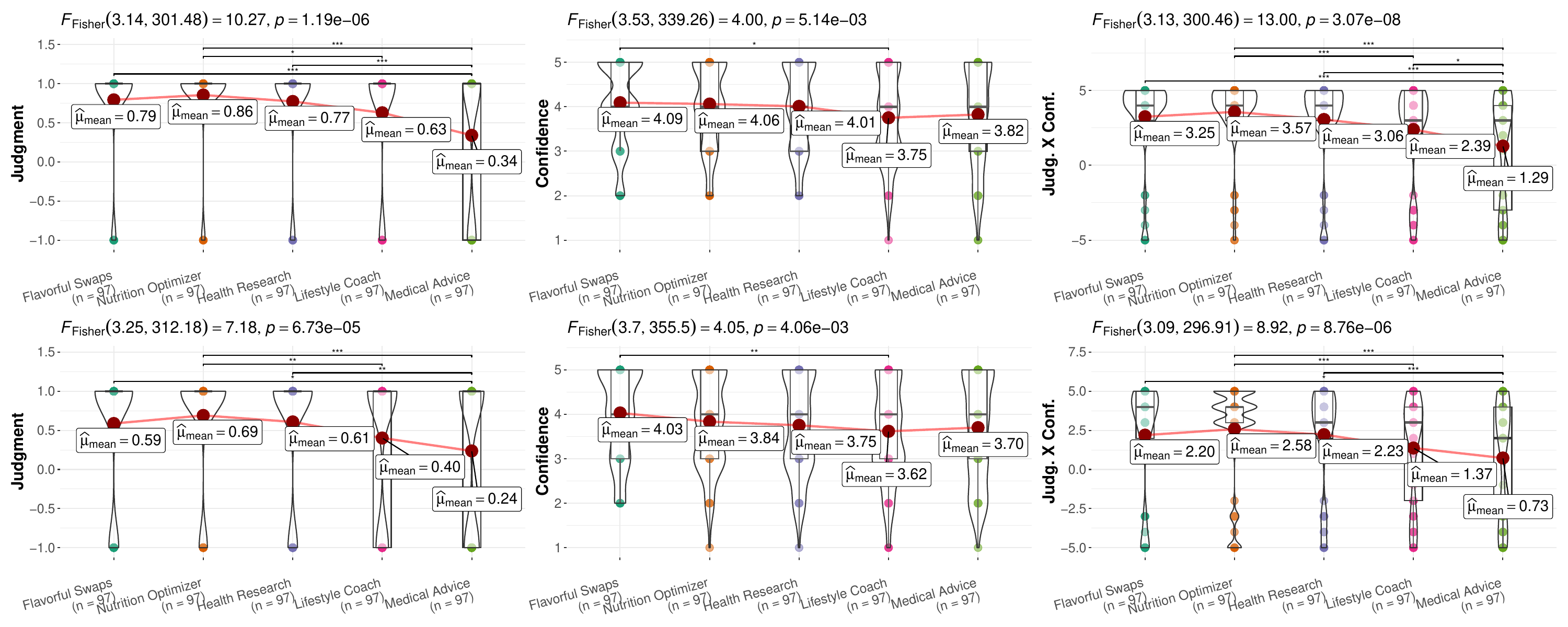}
    \caption{Personal health use case means and distributions of numerically converted Judgment, Confidence, and Judgment$\times$Confidence. First row shows results for development decisions (Q1, Q2) and second row shows results for usage decisions (Q5, Q6). ANOVA results for use cases are shown above each panel. Within subject test was performed using Student's t-test with Holm correction. * denotes following significant p-values: $^{***}p<0.001$; $^{**}p<0.01$; $^{*}p<0.05$. Use case names were shortened.}
    \label{fig:use-case-effect-personal}
\end{figure*}
\paragraph{Personal Health Use Cases}
In personal health use scenarios, Digital Medical Advice AI ($M_{\texttt{DEV}}=0.34, SD_{\texttt{DEV}}=0.95$; $M_{\texttt{USAGE}}=0.24, SD_{\texttt{USAGE}}=0.98$), reflecting high risk usage, consistently had lower acceptance across judgment types, compared to all other use cases. Nutrition Optimizer ($M_{\texttt{DEV}}=0.86, SD_{\texttt{DEV}}=0.92$; $M_{\texttt{USAGE}}=0.69, SD_{\texttt{USAGE}}=0.73$) had the highest mean acceptance across both acceptability judgments.  
Interestingly, unlike the labor-replacement use cases which had slightly higher acceptance for usage, personal use cases had lower acceptance for usage in general compared to development. 

\begin{table}[!hptb]
\centering
\scriptsize
\setlength{\tabcolsep}{0.5pt}
\begin{tabular}{l cc cc}
\hline
 & \multicolumn{2}{c}{\texttt{DEV}($\beta(SE)$)} & \multicolumn{2}{c}{\texttt{USAGE}($\beta(SE)$)}\\
\cmidrule(lr){2-3}\cmidrule(lr){4-5}
 & Judg. & Conf. & Judg. & Conf.\\
\hline
\multicolumn{5}{l}{\textbf{Labor-replacement}} \\
 Coeff. & $\mathbf{-0.08^{**}}(0.03)$ & $-0.00 (0.02)$ & $0.00 (0.03)$ & $-0.03 (0.03)$\\
 $\beta_0$ & $\mathbf{0.43^{***}} (0.10)$ & $\mathbf{3.97^{***}} (0.10)$ & $0.17 (0.10)$ & $\mathbf{4.04^{***}} (0.10)$\\
\addlinespace
\multicolumn{5}{l}{\textbf{Personal Health}} \\
 Coeff. & $\mathbf{-0.11^{***}} (0.02)$ & $\mathbf{-0.08^{***}} (0.02)$ & $\mathbf{-0.10^{***}} (0.02)$ & $\mathbf{-0.09^{***}} (0.02)$\\
 $\beta_0$ & $\mathbf{1.02^{***}} (0.08)$ & $\mathbf{4.20^{***}} (0.10)$ & $\mathbf{0.80^{***}} (0.09)$ & $\mathbf{4.05^{***}} (0.11)$\\
\hline
\multicolumn{5}{l}{\scriptsize{$^{***}p<0.001$; $^{**}p<0.01$; $^{*}p<0.05$}}
\end{tabular}
\caption{Mixed-effects models: use case factor effects (estimate; SE, $\beta_0$ denotes intercept) for judgment and confidence, split by labor-replacement/personal health. Use case variations are numerically coded from 1 (lowest risk/education) to 5 (highest risk/education). Bold indicates $p<0.05$.}
\label{tab:use-case-effect-variations}
\end{table}

\subsubsection{Use Case Variations}
When selecting use cases, we used two underlying variations: entry level of education required for labor-replacement use cases and EU AI risk levels for personal health. As risk levels and required education increased, we observe consistent negative effects on judgments, with personal health use cases showing stronger effects ($\beta_{\texttt{DEV}} = -0.11, p<.001$; $\beta_{\texttt{USAGE}} = -0.10, p < .001$) compared to labor-replacement scenarios ($\beta_{\texttt{DEV}} = -0.08, p<.01$), where only development judgments were significantly associated. 
Confidence ratings showed a small but significant decrease with increasing risk levels in personal health use cases ($\beta_{\texttt{DEV}} = -0.08, p<.001$; $\beta_{\texttt{USAGE}} = -0.09, p<.001$), while labor-replacement use cases showed no significant impact on confidence. 

\subsubsection{Disagreements}
We compare the standard deviation of judgments weighted by confidence to understand possible disagreements and their strength among use cases. Interestingly, the use cases with four highest disagreements in both judgments were all labor-replacement uses in order of Telemarketer ($SD_{\texttt{DEV}}=4.08$; $SD_{\texttt{USAGE}}=4.21$), Elementary School Teacher ($SD_{\texttt{DEV}}=3.99$; $SD_{\texttt{USAGE}}=4.09$), Lawyer ($SD_{\texttt{DEV}}=4.00$; $SD_{\texttt{USAGE}}=3.99$), and Government Eligibility Interviewer AI ($SD_{\texttt{DEV}}=3.96$; $SD_{\texttt{USAGE}}=3.89$). These four use cases were followed by Digital Medical Advice AI ($SD_{\texttt{DEV}}=3.80$, $SD_{\texttt{USAGE}}=3.83$). The use cases with the lowest disagreements were surprisingly Nutrition Optimizer ($SD_{\texttt{DEV}}=2.16$; $SD_{\texttt{USAGE}}=3.03$) followed by IT Support Specialist AI ($SD_{\texttt{DEV}}=3.08$; $SD_{\texttt{USAGE}}=2.65$). 

\paragraph{RQ1 Takeaways} 
Our results show that there are significant variation in judgments based on use case characteristics such as category, EU-defined risk level, and, to some extent, required education level for labor-replacement cases. The uniquely negative response to the Elementary School Teacher AI highlights potential concerns specific to care work. High disagreement in labor-replacement scenarios underscores the need for cautious integration of AI into existing roles. In contrast, the consistent positive judgments for IT Support Specialist AI suggest that not all labor-replacement use cases are viewed equally, indicating the need for nuanced understandings of acceptability. We further examine this variability in \S~\ref{sec:RQ3}. \looseness=-1

\begin{table*}[!h]
\begin{center}
\scriptsize
\begin{tabular}{lcccccc}
\toprule
 & \multicolumn{3}{c}{DEV ($\beta$ (SE))} & \multicolumn{3}{c}{USAGE ($\beta$ (SE))}\\
\cmidrule(lr){2-4}\cmidrule(lr){5-7}
Demographics & Judg. & Conf. & Judg.$\times$Conf. & Judg. & Conf. & Judg.$\times$Conf. \\
\hline
(Intercept) & $\mathbf{0.50^{*}}\;(0.25)$ & $\mathbf{4.08^{***}}\;(0.32)$ & $1.88\;(1.09)$ & $0.51\;(0.28)$ & $\mathbf{3.09^{***}}\;(0.34)$ & $1.34\;(1.23)$ \\
(Intercept)$_{\texttt{Labor}}$ & $0.24\;(.38)$ & $\mathbf{4.59^{***}}\;(.49)$ & $1.16\;(1.66)$ & $0.71\;(.41)$ & $\mathbf{3.74^{***}}\;(.45)$ & $3.09\;(1.74)$ \\
(Intercept)$_{\texttt{Pers}}$ & $0.66\;(.30)$ & $\mathbf{3.76^{***}}\;(.47)$ & $2.33\;(1.34)$ & $0.25\;(.43)$ & $\mathbf{2.61^{***}}\;(.55)$ & $-0.16\;(1.92)$ \\
\textbf{Age} & & & & & & \\
\quad 25-34 & $-0.13\;(0.13)$ & $0.04\;(0.19)$ & $-0.59\;(0.58)$ & $-0.22\;(0.16)$ & $\mathbf{0.41^{*}}\;(0.20)$ & $-0.99\;(0.69)$ \\
\quad 55-64 & $-0.03\;(0.16)$ & $0.32\;(0.23)$ & $-0.14\;(0.69)$ & $0.12\;(0.19)$ & $\mathbf{0.48^{*}}\;(0.24)$ & $0.40\;(0.82)$\\
\quad 25-34$_{\texttt{Pers}}$ & $-0.06\;(.16)$ & $0.36\;(.26)$ & $-0.01\;(.70)$ & $-0.14\;(.23)$ & $\mathbf{0.76^{*}}\;(.30)$ & $-0.10\;(.103)$ \\
\textbf{Race} & & & & & & \\
\quad Asian & $0.17\;(0.10)$ & $\mathbf{-0.37^{**}}\;(0.14)$ & $0.73\;(0.44)$ & $0.10\;(0.12)$ & $\mathbf{-0.33^{*}}\;(0.15)$ & $0.42\;(0.52)$ \\
\quad Black & $0.07\;(0.10)$ & $0.24\;(0.14)$ & $0.49\;(0.43)$ & $0.07\;(0.12)$ & $\mathbf{0.32^{*}}\;(0.15)$ & $0.53\;(0.51)$ \\
\quad Mixed & $0.19\;(0.13)$ & $0.16\;(0.18)$ & $0.85\;(0.56)$ & $-0.13\;(0.15)$ & $\mathbf{0.43^{*}}\;(0.19)$ & $-0.12\;(0.66)$ \\
\quad Asian$_{\texttt{Labor}}$ & $\mathbf{0.40^{**}}\;(.15)$ & $\mathbf{-0.41^{*}}\;(.20)$ & $\mathbf{1.67^{*}}\;(.65)$ & $0.20\;(.16)$ & $\mathbf{-0.44^{*}}\;(.19)$ & $0.59\;(.68)$ \\
\quad Asian$_{\texttt{Pers}}$ & $0.01\;(.12)$ & $\mathbf{-0.54^{**}}\;(.20)$ & $-0.05\;(.56)$ & $0.00\;(.18)$ & $-0.37\;(.24)$ & $0.07\;(.82)$ \\
\quad Black$_{\texttt{Pers}}$ & $0.12\;(.12)$ & $0.35\;(.20)$ & $0.94\;(.55)$ & $0.02\;(.18)$ & $\mathbf{0.59^{*}}\;(.23)$ & $0.65\;(.80)$ \\

\textbf{Gender} & & & & & & \\
\quad Non-male & $\mathbf{-0.29^{***}}\;(0.07)$ & $-0.05\;(0.10)$ & $\mathbf{-1.29^{***}}\;(0.32)$ & $\mathbf{-0.33^{***}}\;(0.09)$ & $0.10\;(0.11)$ & $\mathbf{-1.36^{***}}\;(0.38)$ \\
\quad Non-male$_{\texttt{Labor}}$ & $\mathbf{-0.48^{***}}\;(.11)$ & $0.03\;(.15)$ & $\mathbf{-2.11^{***}}\;(.48)$ & $\mathbf{-0.52^{***}}\;(.12)$ & $0.06\;(.14)$ & $\mathbf{-2.25^{***}}\;(.50)$ \\

\textbf{Political View} & & & & & & \\
\quad Str. liberal & $-0.19\;(0.11)$ & $-0.28\;(0.16)$ & $\mathbf{-1.16^{*}}\;(0.49)$ & $\mathbf{-0.34^{*}}\;(0.13)$ & $0.14\;(0.17)$ & $\mathbf{-1.51^{**}}\;(0.59)$ \\
\quad Str. Liberal$_{\texttt{Labor}}$ & $-0.25\;(.18)$ & $0.02\;(.24)$ & $-1.08\;(.76)$ & $\mathbf{-0.42^{*}}\;(.18)$ & $\mathbf{0.58^{**}}\;(.22)$ & $\mathbf{-1.63^{*}}\;(.79)$ \\
\quad Str. Liberal$_{\texttt{Pers}}$ & $-0.15\;(.16)$ & $\mathbf{-0.53^{*}}\;(.25)$ & $-1.14\;(.70)$ & $-0.18\;(.23)$ & $-0.36\;(.30)$ & $-1.01\;(.102)$ \\
\quad Liberal$_{\texttt{Pers}}$ & $-0.08\;(.11)$ & $\mathbf{-0.52^{**}}\;(.18)$ & $-0.55\;(.50)$ & $-0.07\;(.17)$ & $-0.39\;(.21)$ & $-0.38\;(.74)$ \\

\textbf{Education} & & & & & & \\
\quad Advanced$_{\texttt{Labor}}$ & $\mathbf{0.43^{*}}\;(.19)$ & $-0.48\;(.26)$ & $1.39\;(0.83)$ & $0.31\;(0.20)$ & $-0.40\;(0.24)$ & $1.03\;(0.87)$ \\

\textbf{Employment} & & & & & & \\
\quad 40+ hrs & $\mathbf{0.25^{*}}\;(0.11)$ & $0.07\;(0.16)$ & $\mathbf{1.13^{*}}\;(0.51)$ & $0.09\;(0.14)$ & $0.01\;(0.17)$ & $0.73\;(0.60)$ \\

\textbf{Discrimination} & & & & & & \\
\quad High$_{\texttt{Labor}}$ & $\mathbf{-0.36^{*}}\;(.18)$ & $0.14\;(.25)$ & $\mathbf{-1.79^{*}}\;(.79)$ & $-0.13\;(.19)$ & $0.27\;(.23)$ & $-0.39\;(.82)$ \\
\bottomrule
\multicolumn{7}{l}{\scriptsize{$^{***}p<0.001$; $^{**}p<0.01$; $^{*}p<0.05$}}
\end{tabular}
\caption{Coefficients, Standard Errors, and Significance of Demographic Factors. Models regress decision metrics on demographic factors, with random effects for subjects and use cases. Only significant factors are reported. The intercept represents the dominant demographic group (White, Christian, Male), the lowest natural ordering (18-24, Not Employed), and median values (Moderate, Associate's degree). Subscripts ${\texttt{Labor}}$ and ${\texttt{Pers}}$ indicate labor-replacement and personal health use cases.}
\label{tab:significant-demographics-factor-regression}
\end{center}
\end{table*}

\subsection{RQ2. Impact of Personal Factors on Acceptability Judgment}
\label{sec:RQ2}
\subsubsection{Demographic Factors}
As shown in Table~\ref{tab:significant-demographics-factor-regression}, several demographic factors significantly influenced use case judgments.\footnote{Prior to analysis, we observed that the independent variables in the model had less than 0.5 correlation, except for age 65+ and Retired employment status.} Across both categories of use cases, certain age groups were positively associated with confidence in usage: 25-34 ($\beta_{\texttt{USAGE}}=0.41$, $p<.05$) and 55-64 ($\beta_{\texttt{USAGE}} = 0.48$, $p<.05$). Race also had notable influences; specifically, Asian participants exhibited significantly lower confidence in both development and usage judgments ($\beta_{\texttt{DEV}} = -0.37$, $p<.01$; $\beta_{\texttt{USAGE}} = -0.33$, $p<.05$), particularly in labor-replacement contexts. \looseness=-1

Gender emerged as a crucial determinant, with non-male participants consistently showing negative judgments ($\beta_{\texttt{DEV}} = -0.29$, $p<.001$; $\beta_{\texttt{USAGE}} = -0.33$, $p<.001$), indicating potential discrepancies in perception or experience with AI applications. Liberal views, especially among those identifying as strongly liberal, were associated with negative judgments across both categories of use cases ($\beta_{\texttt{DEV}} = -1.16$, $p<.05$; $\beta_{\texttt{USAGE}} = -1.51$, $p<.01$), suggesting a skeptical stance towards AI's prevalence and role. Employment hours also contributed, with individuals working 40+ hours per week displaying a positive association with development ($\beta_{\texttt{DEV}} = 0.25$, $p<.05$), suggesting more exposure or reliance on AI use cases. High experience of discrimination chronicity was significantly related to lower acceptance of development ($\beta_\texttt{DEV} = -0.36$, $p<.05$). 
See Table~\ref{tab:demographics-anova} in the Appendix for ANOVA results. \looseness=-1

\begin{table*}[!h]
\begin{center}
\scriptsize
\begin{tabular}{l c c c c c c}
\hline
 & \multicolumn{3}{c}{DEV ($\beta$ (SE))} & \multicolumn{3}{c}{USAGE ($\beta$ (SE))}\\
\cmidrule(lr){2-4}\cmidrule(lr){5-7}
AI Literacy & Judg. & Conf. & Judg.$\times$Conf. & Judg. & Conf. & Judg.$\times$Conf. \\
\hline
\multicolumn{7}{l}{\textbf{Labor-replacement}}\\
\quad (Intercept)                   & $0.07\;(0.30)$ & $\mathbf{3.13^{***}}\;(0.33)$ & $-0.49\;(1.27)$ & $-0.50\;(0.32)$ & $\mathbf{3.46^{***}}\;(0.34)$ & $-2.44\;(1.35)$ \\
\quad AI Ethics               & $\mathbf{-0.04^{*}}\;(0.02)$ & $0.02\;(0.02)$ & $-0.15\;(0.08)$ & $-0.00\;(0.02)$ & $-0.01\;(0.02)$ & $-0.06\;(0.08)$ \\
\quad Gen AI Usage Freq.                        & $\mathbf{0.14^{**}}\;(0.04)$ & $-0.00\;(0.05)$ & $\mathbf{0.59^{**}}\;(0.19)$ & $\mathbf{0.18^{***}}\;(0.05)$ & $-0.07\;(0.05)$ & $\mathbf{0.80^{***}}\;(0.20)$ \\
\quad Gen AI Limit. Familiarity                        & $-0.09\;(0.07)$ & $\mathbf{0.20^{*}}\;(0.08)$ & $-0.38\;(0.28)$ & $-0.06\;(0.07)$ & $\mathbf{0.18^{*}}\;(0.08)$ & $-0.38\;(0.30)$ \\
\hline
\multicolumn{7}{l}{\textbf{Personal Health}}\\
\quad (Intercept)                   & $\mathbf{0.87^{***}}\;(0.22)$ & $\mathbf{2.72^{***}}\;(0.40)$ & $\mathbf{2.81^{**}}\;(1.00)$ & $0.49\;(0.30)$ & $\mathbf{2.44^{***}}\;(0.45)$ & $1.17\;(1.30)$ \\
\quad AI Skills                     & $0.00\;(0.01)$ & $\mathbf{0.06^{*}}\;(0.02)$ & $0.07\;(0.06)$ & $0.02\;(0.02)$ & $\mathbf{0.06^{*}}\;(0.03)$ & $\mathbf{0.15^{*}}\;(0.08)$ \\
\quad AI Ethics               & $\mathbf{-0.05^{***}}\;(0.01)$ & $0.01\;(0.03)$ & $\mathbf{-0.23^{***}}\;(0.07)$ & $\mathbf{-0.06^{**}}\;(0.02)$ & $\mathbf{0.07^{*}}\;(0.03)$ & $\mathbf{-0.26^{**}}\;(0.09)$ \\
\quad Gen AI Usage Freq.                        & $\mathbf{0.15^{***}}\;(0.03)$ & $0.06\;(0.06)$ & $\mathbf{0.62^{***}}\;(0.15)$ & $\mathbf{0.19^{***}}\;(0.05)$ & $-0.01\;(0.07)$ & $\mathbf{0.79^{***}}\;(0.20)$ \\
\quad Gen AI Limit. Familiarity                        & $-0.06\;(0.05)$ & $0.00\;(0.09)$ & $-0.25\;(0.21)$ & $-0.10\;(0.06)$ & $-0.10\;(0.10)$ & $-0.50\;(0.28)$ \\
\hline
\multicolumn{7}{l}{\scriptsize{$^{***}p<0.001$; $^{**}p<0.01$; $^{*}p<0.05$}}
\end{tabular}
\caption{Coefficients, Standard Errors, and Significance of AI Literacy Factors. Models regress decision metrics on AI literacy factors, with random effects for subjects and use cases. Only significant factors are reported.}
\label{tab:ai-literacy-effects}
\end{center}
\end{table*}

\subsubsection{AI Literacy}
We identified a correlation greater than 0.5 among three AI literacy aspects: awareness, usage, and evaluation. To avoid multicollinearity, we aggregated them into a single factor, AI Skills. As shown in Table~\ref{tab:ai-literacy-effects}, understanding of AI Ethics was associated with lower acceptability for both personal health ($\beta_{\texttt{DEV}}=-0.05$, $p<.001$; $\beta_{\texttt{USAGE}}=-0.23$, $p<.001$) and labor-replacement ($\beta_{\texttt{DEV}} = -0.04,p<.05$). However, across both categories, high Generative AI Usage Frequency resulted in higher acceptance (Labor-replacement -- $\beta_{\texttt{DEV}}=0.14$, $p<.01$; $\beta_{\texttt{USAGE}}=0.18$, $p<.001$, Personal Health -- $\beta_{\texttt{DEV}}=0.15$, $p<.001$; $\beta_{\texttt{USAGE}}=0.19$, $p<.001$). Notably, for personal health use cases, AI Skills was positively associated with confidence of judgments ($\beta_{\texttt{DEV,USAGE}}=0.06$, $p<.05$), while Generative AI Limitation Familiarity was positively associated with confidence for labor-replacement usage ($\beta_{\texttt{DEV}}=0.20$, $p<.05$; $\beta_{\texttt{USAGE}}=0.18$, $p<.05$ ). 
\paragraph{RQ2 Takeaways} 
Our findings highlight the significant role of lived experiences and backgrounds, such as age, race, gender, political view, employment, and discrimination experience, in shaping perceptions and attitudes toward AI technologies. Moreover, our results indicate that different understandings of and experiences with AI can impact judgments of acceptability, corroborating previous findings \citep{kramer2018when}.

\begin{figure*}[!h]
    \centering
    \includegraphics[width=\linewidth]{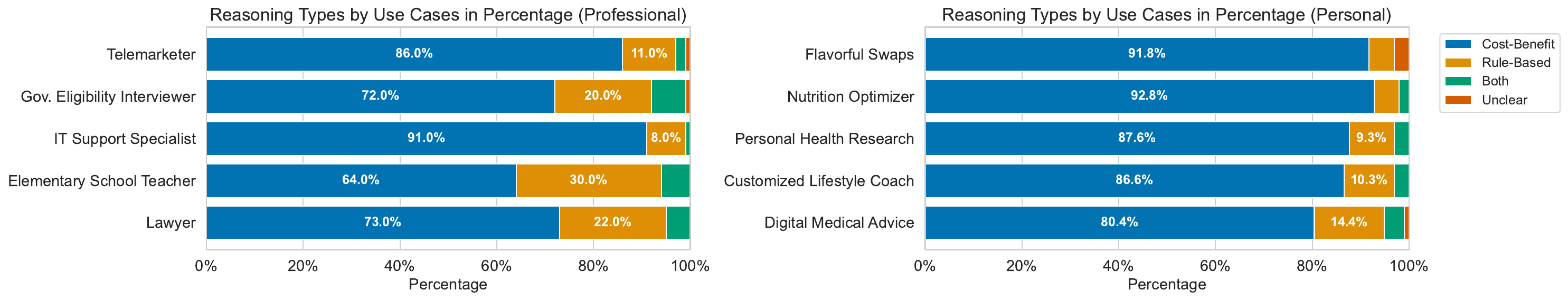}
    \caption{Percentage of reasoning types (cost-benefit and rule-based) by use cases in participant provided rationales (Q3, Q4).}
    \label{fig:reasoning-types-percentage}
\end{figure*}
\begin{figure*}[!h]
    \centering
    \includegraphics[width=\linewidth]{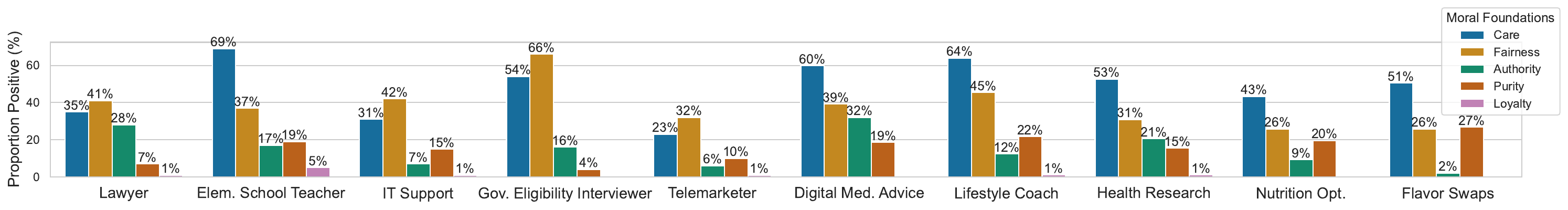}
    \caption{Proportion (\%) of presence of moral foundations in participant's rationale responses (Q3, Q4) aggregated by use case.}
    \label{fig:moral_values_proportions}
\end{figure*}
\begin{table*}[!hptb]
\centering
\scriptsize
\setlength{\tabcolsep}{3pt}
\renewcommand{\arraystretch}{0.9}
\begin{tabular}{llcccccc}
\toprule
    &  & \multicolumn{3}{c}{DEV ($\beta$ (SE))} & \multicolumn{3}{c}{USAGE ($\beta$ (SE))} \\
    \cmidrule(lr){3-5} \cmidrule(lr){6-8}
    &  & Judg. & Conf. & Judg.$\times$Conf. & Judg. & Conf. & Judg.$\times$Conf. \\
\midrule
\multicolumn{8}{l}{\textbf{Reasoning Type}} \\
    & Cost-benefit
        & $\mathbf{0.32^{**}\;(0.10)}$ & $0.15\;(0.15)$ & $\mathbf{1.31^{***}\;(0.39)}$
        & $\mathbf{-0.10^{*}\;(0.04)}$ & $\mathbf{0.32^{*}\;(0.16)}$ & $\mathbf{2.40^{***}\;(0.52)}$ \\
    & Rule-based
        & $\mathbf{-0.46^{***}\;(0.09)}$ & $\mathbf{0.43^{**}\;(0.14)}$ & $\mathbf{-1.61^{***}\;(0.35)}$
        & $-0.06\;(0.04)$ & $0.25\;(0.14)$ & $-0.88\;(0.48)$ \\
\multicolumn{8}{l}{\textbf{Moral Value}} \\
    & Care
        & $0.05\;(0.04)$ & $-0.06\;(0.06)$ & $0.15\;(0.15)$
        & $-0.00\;(0.02)$ & $\mathbf{-0.15^{*}\;(0.06)}$ & $-0.37\;(0.21)$ \\
    & Fairness
        & $\mathbf{0.14^{***}\;(0.04)}$ & $\mathbf{-0.17^{**}\;(0.06)}$ & $\mathbf{0.53^{***}\;(0.16)}$
        & $0.00\;(0.02)$ & $-0.08\;(0.07)$ & $\mathbf{0.71^{**}\;(0.22)}$ \\
    & Authority
        & $\mathbf{-0.20^{***}\;(0.05)}$ & $-0.06\;(0.08)$ & $\mathbf{-0.77^{***}\;(0.20)}$
        & $-0.00\;(0.02)$ & $-0.08\;(0.08)$ & $-0.54\;(0.28)$ \\
\multicolumn{8}{l}{\textbf{Switching Condition}} \\
    & Usage
        & $\mathbf{0.12^{***}\;(0.01)}$ & $0.02\;(0.01)$ & $\mathbf{0.56^{***}\;(0.02)}$
        & $\mathbf{0.24^{***}\;(0.00)}$ & $\mathbf{-0.04^{***}\;(0.01)}$ &   \\
    & Societal Impact
        & $-0.08\;(0.04)$ & $0.00\;(0.07)$ & $\mathbf{-0.39^{*}\;(0.17)}$
        & $0.03\;(0.02)$ & $-0.12\;(0.07)$ & $\mathbf{-0.74^{**}\;(0.23)}$ \\
\midrule
    & (Intercept)
        & $0.09\;(0.11)$
        & $\mathbf{3.88^{***}\;(0.16)}$
        & $0.18\;(0.44)$
        & $0.16^{***}\;(0.04)$
        & $\mathbf{3.81^{***}\;(0.17)}$
        & $-0.53\;(0.64)$ \\
\bottomrule
\multicolumn{8}{l}{\scriptsize{$^{***}p{<}0.001$; $^{**}p{<}0.01$; $^{*}p{<}0.05$}}
\end{tabular}
\caption{Coefficients (SE) and significance of rationale factors. All factors coded as binary. Only showing significant results.} 
\label{tab:reasoning-factors-judgment-regression}
\end{table*}

\subsection{RQ3. Factors in Participant Rationale}
\label{sec:RQ3}
To deepen our analysis of use case acceptability, we examined open-text rationales (Q3) and decision-switching conditions (Q4) related to development judgments (Q1).
\subsubsection{Decision-making Types}
As we defined in \S~\ref{ssec:text-analysis}, we focus on two distinct reasoning types for decision-making: cost-benefit reasoning, which emphasizes outcomes (e.g., \textit{``it gives more people access to medical advice and treatment''}, P365), and rule-based reasoning, which reflects values inherent in the action itself (e.g., should not be developed because \textit{``human interaction is better''}, P249). 
As shown in Figure~\ref{fig:reasoning-types-percentage}, generally, participants used more cost-benefit reasoning, especially for the IT Support Specialist (91.0\%) and Nutrition Optimizer (92.8\%) use cases. This result is particularly interesting as we observed these two use cases to have the lowest disagreement (see \S~\ref{sec:RQ1}), suggesting unified reasoning type may lead to more consistent judgments. On the other hand, rationales for Elementary School Teacher AI contained most percentage of rule-based reasoning (30.0\%) followed by Lawyer AI (22.0\%), both of which were the two use cases with lowest development acceptability.\looseness=-1 

Through further analysis of the influence of rationale factors on judgments using a mixed-effects model (Table~\ref{tab:reasoning-factors-judgment-regression}), we found that use of cost-benefit reasoning in rationale is positively associated with development acceptability ($\beta_{\texttt{DEV}} = 0.32, p<.01$), while rule-based reasoning is negatively associated ($\beta_{\texttt{DEV}} = -0.46, p<.001$). Interestingly, for usage, cost-benefit reasoning was negatively associated with acceptability ($\beta_{\texttt{USAGE}}=-0.10,p<.05$) but positively for judgment weighted by confidence ($\beta_{\texttt{USAGE}}=2.40,p<.001$). This suggests that, although cost-benefit considerations may reduce usage acceptance, they boost confidence when judgments are favorable. 

The significant negative association of rule-based reasoning with judgments indicate that for certain contexts, the inherent action of using AI is viewed negatively. Participants cited diverse concerns related to AI itself (e.g., \textit{``because it would not have human sympathy''}, P16), human needs (e.g., \textit{``Humans need human interactions in order to learn properly''}, P44), societal impact (e.g., \textit{``Overreliance on AI ...''}, P132), and morality (e.g., \textit{``Having artificial intelligence try and sell you things is immoral''}, P13). These results suggest that AI's acceptability heavily lies in its positive outcomes but can be outweighed by established rules. 

\subsubsection{Moral Foundations}
Beyond reasoning types, we explored moral foundations to provide insights into what values are relevant for AI use case decisions (Figure ~\ref{fig:moral_values_proportions}). For example, P12 responded that Elementary School Teacher AI use case should be developed because it \textit{``could give elementary schooling to children who are bed ridden...''}, which was annotated with both values of Care (focusing on the well-being and nurturing of bed-ridden children) and Fairness (focusing on fair access to education).
Upon analysis, we observe that Care (i.e., dislike of pain of others, feelings of empathy and compassion toward others) was the most prevalent moral foundation in participants' rationales across the use cases in both categories (48\%). Interestingly, Care could be invoked in both positive and negative regards for AI, as conveyed by P385 who noted that Customized Lifestyle Coach AI should be developed because \textit{``it may help improve some people's health''} but would change their decision if \textit{``it caused harm to even one person.''}

Although Care was the most dominant moral foundation overall, Fairness emerged more prominently in context-specific evaluations such as Lawyer AI (41\%) and Government Eligibility Interviewer AI (66\%). These results could be due to the characteristics of the use cases, such as their main purpose and function; as noted by P88, Government Eligibility Interviewer should be developed because \textit{``it might be less biased and therefore more fair in it decisions (sic)''}. Authority was most apparent in participant rationales for Lawyer AI (28\%) and Purity for Flavorful Swaps (27\%). Moreover, Fairness in rationales had positive associations with acceptance ($\beta_{\texttt{DEV}} = 0.53, p<.001$, $\beta_{\texttt{USAGE}} = 0.71, p<.01$; judgment weighted by confidence; see Table~\ref{tab:reasoning-factors-judgment-regression}). 

\begin{figure*}[!h]
    \centering
    \includegraphics[width=\linewidth]{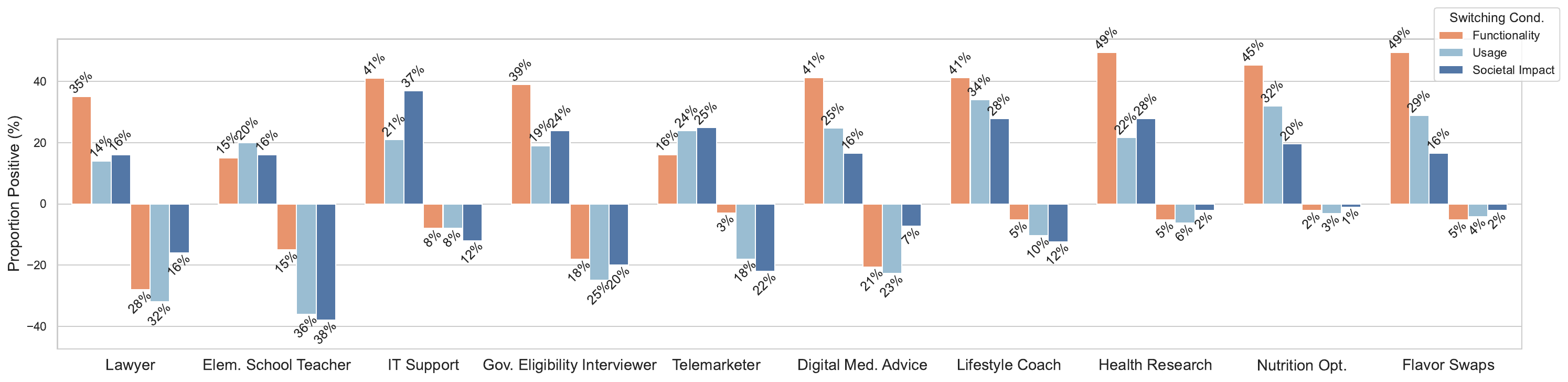}
    \caption{Proportion (\%) of presence of switching conditions (Functionality, Usage, Societal Impact) mentioned in participant's switching conditions (Q4) aggregated by use case and divided into positive and negative development acceptability. The total proportion of the switching condition by use case is the sum of both positive and negative bars.}
    \label{fig:switching_conditions_proportions}
\end{figure*}
\subsubsection{Switching Conditions}
We further explored the flexibility of participants' judgments to understand possible mitigation of disagreements through criteria for switching their decisions (Figure~\ref{fig:switching_conditions_proportions}). Functionality (53\%; e.g., Medical Adivce AI should not be developed if it \textit{``consistently or had a high percentage of failure to diagnose correctly.''}, P373) was the most commonly noted condition for switching decisions in both directions (positive to negative and vice versa). This was followed by Usage (40\%; e.g., Government Eligibility Interviewer should be developed if \textit{``it was only used to read and screen applications but not for making decisions''}), Societal Impact (36\%; e.g., Lawyer AI should not be developed if \textit{``it puts too many human lawyers out of work''}, P97), and Not Applicable (7\%; e.g., will not change decision). 

Interestingly, for labor-replacement use cases Societal Impact (45\%) was more frequently mentioned as switching conditions followed by Functionality (43\%), whereas Functionality (53\%) was more frequently mentioned than Societal Impact (37\%) for personal use cases. Frequency of Societal Impact in labor-replacement use cases could be closely linked to concerns of labor replacement: as described by P296, if Elementary School Teacher AI \textit{``was to replace teachers with the ai to save money''}, they would switch their decision from positive to negative. Moreover, for all use cases except Elementary School Teacher AI, participants tended to switch from positive to negative decisions for lack of functionality reasons. In contrast, for Elementary School Teacher AI, the most common shift towards acceptability when the use case showed a positive societal impact (38\%). As shown in Table~\ref{tab:reasoning-factors-judgment-regression}, mentions of Societal Impact as conditions to switch decisions were more negatively associated with judgments ($\beta_{\texttt{DEV}} = -0.39, \, p<.05$, $\beta_{\texttt{USAGE}} = -0.74, \, p<.05$; judgment $\times$ confidence). However, emphasis on Usage ($\beta_{\texttt{DEV}} = 0.12, \, p<.001$) as a condition to reverse their decisions was positively associated. \looseness=-1

\paragraph{RQ3 Takeaways}
Reasoning types varied by context, with rule-based reasoning more common in contested use cases and negatively associated with acceptability, while cost-benefit reasoning showed a positive association. The moral foundation of Care was especially salient, highlighting its importance in AI judgments. When explaining what might change their decisions, participants most frequently cited Functionality for personal health use cases and Societal Impact for labor-replacement, underscoring the context-dependent nature of these concerns, especially to be considered when mitigating disagreements.
\section{Conclusion and Discussion}

We conducted a study to understand how and why laypeople perceive various AI use cases as acceptable or not. To achieve this, we developed a survey that gathered judgments and reasoning processes from 197 participants who were demographically diverse and had varying levels of experience with AI. Participants were asked to provide their judgments on the acceptability of AI use cases, along with rationales for their decisions (e.g., ``Should / Should not be developed, because...") and conditions that might change their decisions (e.g., ``I would switch my decision if..."). The survey covered ten different AI use cases, spanning both personal and professional domains, and included varying levels of risk. Our findings revealed significant variation in the acceptability judgments and reasoning factors based on the domain, risk level, and participants' attributes, such as AI literacy and gender. We discuss the implications of these findings below. \looseness=-1

\paragraph{Use Case Perceptions and Disagreements}
In our study, we explored the varying acceptability of AI across different use cases. Generally, acceptance was lower in scenarios with higher educational requirements and greater EU AI risk levels. Professional use cases displayed more variability, notably with Elementary School Teacher AI, which was uniquely unacceptable. This underscores the necessity for further research into how AI should be developed and integrated, as well as what skills it should have, particularly in fields where empathy and care are crucial \citep{wu2024care,kawakami2024ai,borg2024required}. In addition, prior research have also highlighted how AI practitioners desire understanding lay people's perception on AI fairness in specific use cases \citep{sonboli2021fairness, deng2022exploring, smith2023scoping, deng2023understanding}. Drawing from prior HCI and AI research \citep{deng2025weaudit, lee2019webuildai, cheng2019explaining}, future researchers and practitioners should explore how to meaningfully connect lay people's use case perceptions with AI developers' workflows. \looseness=-1

While prior research has emphasized understanding AI consequences \citep{kieslich2024myfuture} and providing tools and processes to uncover impact \citep{wang2024farsight,buccinca2023aha, deng2024supporting}, our findings reveal a greater presence of rule-based reasoning in contentious use cases, suggesting a need for diverse approaches to understanding AI beyond mere consequence anticipation. Moreover, while care was generally predominant, we observed that fairness gained prominence in Lawyer AI and Government Eligibility Interviewer AI. This variability underscores the importance of considering values in AI evaluation and training \citep{barocas2021evaluation,bhardwaj2024machine}, rather than solely emphasizing functionality, which is the current trend in AI research \citep{birhane2022values}. Additionally, societal impact considerations were more evident in unacceptable use cases, emphasizing the necessity for implementing safety guardrails when deploying AI with significant social implications \citep{solaiman2023genairelease}. \looseness=-1

\paragraph{Demographics and AI Literacy}
In line with prior work \cite{kingsley2024investigating, mun2024participaidemocraticsurveyingframework}, our results highlighted significant differences among demographic groups and perceived acceptance of use cases, especially for professional use (\S\ref{sec:RQ2}). Non-majority demographic groups, especially non-male gender groups, found both personal and professional use cases less acceptable. Those experiencing high discrimination chronicity also found professional use less acceptable. 
Our findings offer empirical insights for future research on AI integration in workplaces, where marginalized workers’ agency, income, and well-being are disproportionately impacted \citep{ming2024labor,alcover2021aging}. 

Furthermore, our work highlighted a potential polarization on perceptions of AI among workers as those with 40+ hours employment and those who had advanced degrees were more positive towards AI use cases, suggesting that the relationship stakeholders have to AI and jobs might influence acceptability. This concern was expressed by a participant who opposed the development of Telemarketer AI, stating that it \textit{“overlaps with my industry, and hence serves as a threat to my job security”} (P35). Thus, our results corroborate the need to further explore methods to include diverse workers and various stakeholders into the discussion of workplace AI integration and development \citep{fox2020worker,cheon2023bigtechwork}. We also found that frequent AI usage increased acceptance, while understanding AI ethics and limitations decreased acceptance. This suggests that balanced AI awareness and education, encompassing usage, skills, and ethics, could guide and improve decision-making \citep{raji2021ethics}, e.g., through educational interventions targeting AI skills and ethical implication literacy \citep[e.g.,][]{wong2021timelines, shen2021value}. \looseness=-1
\paragraph{Rationales}
Through analyzing participants' rationales, we observed an interesting pattern: use cases with less disagreement tended to elicit more cost-benefit (utilitarian) reasoning, while those with greater disagreement showed more rule-based (deontological) reasoning (\S\ref{sec:RQ3}). This suggests participants may apply different valuation frameworks, leading to diverging judgments, and highlights that some use cases raise concerns beyond simple utilitarian considerations. Our results thus underscore crucial elements of participants decision making pattern when only assessing impact as many prior works have done. Building upon our empirical findings, future work could develop diverse open-ended analysis for eliciting deliberations as well as tools and interventions using specific types of acceptability reasoning such as rule and value based \cite{sorensen2024valueKaleidoscope} or cost-benefit analyses \cite{li2024safetyanalyst}. \looseness=-1

However, as our study was limited to the two reasoning type categories, expanding this analysis would be essential for future work including finding ways to classify what features people are considering in their decisions, how the weights on those features impact what kind of decision-making strategy they will use, and whether there are other ways to understand their decision strategies beyond our current classification. Future research can build upon these further understandings to guide policy making and consensus building. For example, future work could explore how group discussions, beyond individual surveys, shape communities' collective understanding of AI impacts  \citep[e.g.,][]{kuo2024policycraft, lee2019webuildai,devos2022toward, gordon2022jury,zhang2023deliberating}. \looseness=-1

\bibliography{custom}
\appendix

\section{Additional Survey Details}
\label{app:survey-details}

\subsection{Survey Questions}\label{sssec:a-questions}
In our survey we ask participants to read one or more descriptions of AI use cases and to make two judgments: 1) ``Do you think a technology like this should exist?'' (Q1) and 2) ``If the <\textit{use case}> exists, would you use its services?'' (Q5). A question to indicate their level of confidence is asked following each question (Q5, Q6). The participants are asked to both elaborate on their decisions (Q3) and specify the conditions under which they would switch their decisions (Q4). The detailed wordings for the questions are shown in Table \ref{app:part-1-questions}.

Following the main use case questions in both the main and second study, we also asked participants questions about their demographics and literacy levels in AI, and the questions can be found in Table \ref{app:demographic-questions} and \ref{app:ai-literacy-questions} respectively. 

Lastly, while not included in the main text, we asked participants 3 questionnaires about decision making styles to explore the relationship between several decision making styles and the actual decisions of the participants. These included: (1) Oxford Utilitarianism Scale, (2) Toronto Empathy Questionnaire and (3) Moral Foundations Questionnaire. The decision making style questions can be found in Table \ref{app:util-questions}, \ref{app:empathy-questions} and \ref{app:mfq} respectively.

\begin{table*}[!hbpt]
    \footnotesize
    \centering
    \begin{tabularx}{\linewidth}{l|X|l}
    \toprule
         Question ID & Question & Answer Type \\
         \midrule
         \multicolumn{3}{l}{AI Perception Question (Before)}\\
         \midrule
         AI Perception Before & Overall, how does the growing presence of artificial intelligence (AI) in daily life and society make you feel? & 5 Point Likert Scale\\
         \midrule
         \multicolumn{3}{l}{Part 1 Specific Questions}\\
         \midrule
         UCX-1 & Do you think a technology like this should be developed? & Yes/No\\
         UCX-2 & How confident are you in your above answer? & 5 Point Likert Scale \\
         UCX-3Y & Please complete the following: [Use Case] should be developed because... & Text \\
         UCX-3N & Please complete the following: [Use Case] should not be developed because... & Text \\
         UCX-4Y & Under what circumstances would you switch your decision from [UCX-2 Answer] should be developed to should not be developed? & Text \\
         UCX-4N & Under what circumstances would you switch your decision from [UCX-2 Answer] should not be developed to should be developed? & Text \\ 
         UCX-5 & If [Use Case] exists, would you ever use its services (answer yes, even if you think you would use it very infrequently)? & Yes/No \\
         UCX-6 & How confident are you in your above answer? & 5 Point Likert Scale \\
         \midrule
         \multicolumn{3}{l}{AI Perception Question (After)}\\
         \midrule
         AI Perception After & Before we continue, we’d like to get your thoughts on AI one more time. Overall, how does the growing presence of artificial intelligence (AI) in daily life and society make you feel? & 5 Point Likert Scale \\
    \bottomrule
    \end{tabularx}
    \caption{Main Study Specific Question, the "X" in Question IDs is a placeholder for the use case number, which ranges from 1 to 5, for the 5 use cases in the jobs and personal use cases respectively.}
    \label{app:part-1-questions}
\end{table*}
\begin{table}[!hbpt]
    \footnotesize
    \centering
    \begin{tabularx}{\linewidth}{l|X}
    \toprule
         Question ID & Question  \\
         \midrule
         D-Q1 & How old are you? \\
         D-Q2 & Choose one or more races that you consider yourself to be \\
         D-Q3 & Do you identify as transgender? \\
         D-Q4 & How would you describe your gender identity? \\
         D-Q5 & How would you describe your sexual orientation? \\
         D-Q6 & What is your present religion or religiosity, if any? \\
         D-Q7 & In general, would you describe your political views as… \\
         D-Q8 & What is the highest level of education you have completed? \\
         D-Q9 & In which country have you lived in the longest? \\
         D-Q10 & What other countries have you lived in for at least 6 months? \\
         D-Q11 & Which of the following categories best describe your employment status? \\
         D-Q12 & How would you describe the industry your job would be in? (Select all that apply) \\
         D-Q13 & Do you identify with any minority, disadvantaged, demographic, or other specific groups? If so, which one(s)? (E.g., racial, gender identity, sexuality, disability, immigrant, veteran, etc.); use commas to separate groups. \\
         D-Q14 & (Optional) What are some things that you are most concerned about lately? \\
         $D-Q15_1$ & In your day-to-day life how often have any of the following things happened to you? You are treated with less courtesy or respect than other people \\
         $D-Q15_2$ & In your day-to-day life how often have any of the following things happened to you? You receive poorer service than other people at restaurants or stores \\
         $D-Q15_3$ & In your day-to-day life how often have any of the following things happened to you? People act as if they think you are not smart\\
         $D-Q15_4$ & In your day-to-day life how often have any of the following things happened to you? People act as if they are afraid of you\\
         $D-Q15_5$ & In your day-to-day life how often have any of the following things happened to you? You are threatened or harassed\\
         D-Q16 & If the answer to Q15 is ''A few times a year'' or more frequently to at least one of the statements, what do you think is the main reason for these experiences? (Select all that apply) \\
    \bottomrule
    \end{tabularx}
    \caption{Demographic Questions}
    \label{app:demographic-questions}
\end{table}
\begin{table}[!hbpt]
    \footnotesize
    \centering
    \begin{tabularx}{\linewidth}{l|X}
    \toprule
         Question ID & Question\\
         \midrule
         AI-Q1 & I can identify the AI technology employed in the applications and products I use. \\
         AI-Q2 & I can skillfully use AI applications or products to help me with my daily work. \\
         AI-Q3 & I can choose the most appropriate AI application or product from a variety for a particular task. \\
         AI-Q4 & I always comply with ethical principles when using AI applications or products. \\
         AI-Q5 & I am never alert to privacy and information security issues when using AI applications or products. \\
         AI-Q6 & I am always alert to the abuse of AI technology. \\
         AI-Q7 & How frequently do you use generative AI (i.e., artificial intelligence that is capable of producing high quality texts, images, etc. in response to prompts) products such as ChatGPT, Bard, DALL·E 2, Claude, etc.? \\
         AI-Q8 & How familiar are you with limitations and shortcomings of generative AI? \\
    \bottomrule
    \end{tabularx}
    \caption{AI Literacy Questions. The questions are on a 7 point likert scale ranging from Strongly disagree to Neutral to Strongly agree}
    \label{app:ai-literacy-questions}
\end{table}
\begin{table}[!hbpt]
    \footnotesize
    \centering
    \begin{tabularx}{\linewidth}{l|X}
    \toprule
         Question ID & Question \\
         \midrule
         Util1 & If the only way to save another person’s life during an emergency is to sacrifice one’s own leg, then one is morally required to make this sacrifice. \\
         Util2 & It is morally right to harm an innocent person if harming them is a necessary means to helping several other innocent people. \\
         Util3 & From a moral point of view, we should feel obliged to give one of our kidneys to a person with kidney failure since we don’t need two kidneys to survive, but really only one to be healthy. \\
         Util4 & If the only way to ensure the overall well-being and happiness of the people is through the use of political oppression for a short, limited period, then political oppression should be used. \\
         Util5 & From a moral perspective, people should care about the well-being of all human beings on the planet equally; they should not favor the well-being of people who are especially close to them either physically or emotionally \\
         Util6 & It is permissible to torture an innocent person if this would be necessary to provide information to prevent a bomb going off that would kill hundreds of people. \\
         Util7 & It is just as wrong to fail to help someone as it is to actively harm them yourself. \\
         Util8 & Sometimes it is morally necessary for innocent people to die as collateral damage if more people are saved overall. \\
         Util9 & It is morally wrong to keep money that one doesn’t really need if one can donate it to causes that provide effective help to those who will benefit a great deal. \\
    \bottomrule
    \end{tabularx}
    \caption{Utilitarianism Questions. The questions are on a 7-point likert scale ranging from 1 (Strongly Disagree) to 7 (Strongly Agree)}
    \label{app:util-questions}
\end{table}
\begin{table}[!hbpt]
    \footnotesize
    \centering
    \begin{tabularx}{\linewidth}{l|X}
    \toprule
         Question ID & Question \\
         \midrule
         Empathy1 & When someone else is feeling excited, I tend to get excited too. \\
         Empathy2 & Other people’s misfortunes do not disturb me a great deal. \\
         Empathy3 & It upsets me to see someone being treated disrespectfully. \\
         Empathy4 & I remain unaffected when someone close to me is happy. \\
         Empathy5 & I enjoy making other people feel better. \\
         Empathy6 & I have tender, concerned feelings for people less fortunate than me. \\
         Empathy7 & When a friend starts to talk about his/her problems, I try to steer the conversation towards something else. \\
         Empathy8 & I can tell when others are sad even when they do not say anything. \\
         Empathy9 & I find that I am “in tune” with other people’s moods. \\
         Empathy10 & I do not feel sympathy for people who cause their own serious illnesses. \\
         Empathy11 & I become irritated when someone cries. \\
         Empathy12 & I am not really interested in how other people feel. \\
         Empathy13 & I get a strong urge to help when I see someone who is upset. \\
         Empathy14 & When I see someone being treated unfairly, I do not feel very much pity for them. \\
         Empathy15 & I find it silly for people to cry out of happiness. \\
         Empathy16 & When I see someone being taken advantage of, I feel kind of protective towards him/her. \\
    \bottomrule
    \end{tabularx}
    \caption{Empathy Questions. The questions are on a 5 point likert scale ranging from Never to Always.}
    \label{app:empathy-questions}
\end{table}
\begin{table}[!hbpt]
    \footnotesize
    \centering
    \begin{tabularx}{\linewidth}{l|X}
    \toprule
         Question ID & Question \\
         \midrule
         \multicolumn{2}{X}{Moral Foundation Questionnaire (First Half)}\\
         \midrule
         \multicolumn{2}{X}{When you decide whether something is right or wrong, to what extent is the following consideration relevant to your thinking?}\\
         \midrule
         MFQ 1 & Whether or not someone suffered emotionally \\
         MFQ 2 & Whether or not some people were treated differently than others\\
         MFQ 3 & Whether or not someone’s action showed love for his or her country \\
         MFQ 4 & Whether or not someone showed a lack of respect for authority \\
         MFQ 5 & Whether or not someone violated standards of purity and decency \\
         MFQ 6 & Whether or not someone was good at math \\
         MFQ 7 & Whether or not someone cared for someone weak or vulnerable \\
         MFQ 8 & Whether or not someone acted unfairly \\
         MFQ 9 & Whether or not someone did something to betray his or her group \\
         MFQ 10 & Whether or not someone conformed to the traditions of society \\
         MFQ 11 & Whether or not someone did something disgusting \\
         \midrule
         \multicolumn{2}{l}{Moral Foundation Questionnaire (Second Half)}\\
         \midrule
         MFQ 12 & Compassion for those who are suffering is the most crucial virtue.\\
         MFQ 13 & When the government makes laws, the number one principle should be ensuring that everyone is treated fairly.\\
         MFQ 14 & I am proud of my country’s history.\\
         MFQ 15 & Respect for authority is something all children need to learn.\\
         MFQ 16 & People should not do things that are disgusting, even if no one is harmed.\\
         MFQ 17 & It is better to do good than to do bad.\\
         MFQ 18 & One of the worst things a person could do is hurt a defenseless animal.\\
         MFQ 19 & Justice is the most important requirement for a society.\\
         MFQ 20 & People should be loyal to their family members, even when they have done something wrong.\\
         MFQ 21 & Men and women each have different roles to play in society.\\
         MFQ 22 & I would call some acts wrong on the grounds that they are unnatural.\\
    \bottomrule
    \end{tabularx}
    \caption{Moral Foundation Questionnaire: 20 Questions. The first part of the questionnaire consists of 11 questions on a 6-point likert scale ranging from 0 (Not At All Relevant) to 5 (Extremely Relevant). The second part of the questionnaire consists of 11 questions on a 6-point likert scale ranging from 0 (Strongly Disagree) to 5 (Strongly Agree). Note: Questions MFQ 6 and 17 are meant to catch participants that are not answering the questionnaire properly and are not included in the MFQ score calculation.}
    \label{app:mfq}
\end{table}

\subsection{Use Case Descriptions}\label{app:use-case-descriptions}
See Table\ref{tab:use-case-description-long-labor} for labor replacement use cases as shown to the participants and Table\ref{tab:use-case-descriptions-long-personal-health} for personal health use cases. 
\begin{table*}[h!]
\footnotesize
\centering
\begin{tabularx}{\textwidth}{l|p{0.2\linewidth}|X}
\hline
\textbf{Use Case} & \textbf{Education Level} & \textbf{Description} \\ \hline

Lawyer AI &
High: Doctoral or Professional Degree, Master’s Degree &
Imagine an AI system that performs the tasks of a Lawyer digitally. The AI system advises and represents clients on digital legal proceedings or transactions. The system can read, write, and talk fluently. It has expert-level knowledge, can browse the internet, and can process and read massive amounts of data. We will refer to the AI system as Lawyer AI in the rest of the survey. \\
\hline

Elementary School Teacher AI &
Middle High: Bachelor Degree, Associate Degree &
Imagine an AI system that performs the tasks of an Elementary School Teacher digitally. The AI system teaches academic skills to students at the elementary school level through online interactions. The system can read, write, and talk fluently. It has expert-level knowledge, can browse the internet, and can process and read massive amounts of data. We will refer to the AI system as Elementary School Teacher AI in the rest of the survey. \\
\hline

IT Support Specialist AI &
Middle: Some College No Degree, Postsecondary Nondegree Award &
Imagine an AI system that performs the tasks of an IT Support Specialist digitally. The AI system maintains computer networks and provides technical help to computer users. The system can read, write, and talk fluently. It has expert-level knowledge, can browse the internet, and can process and read massive amounts of data. We will refer to the AI system as IT Support Specialist AI in the rest of the survey. \\
\hline

Government Eligibility Interviewer AI &
Middle Low: High School Diploma or Equivalent &
Imagine an AI system that performs the tasks of an Eligibility Interviewer for Government Programs digitally. This AI system determines eligibility of persons applying to receive assistance from government programs and agency resources, such as welfare, unemployment benefits, social security, and public housing. The system can read, write, and talk fluently. It has expert-level knowledge, can browse the internet, and can process and read massive amounts of data. We will refer to the AI system as Government Eligibility Interviewer AI in the rest of the survey. \\
\hline

Telemarketer AI &
Low: No Formal Educational Credential &
Imagine an AI system that performs the tasks of a Telemarketer digitally. The AI system solicits donations or orders for goods or services over the telephone. The system can read, write, and talk fluently. It has expert-level knowledge, can browse the internet, and can process and read massive amounts of data. We will refer to the AI system as Telemarketer AI in the rest of the survey. \\
\hline

\end{tabularx}
\caption{Labor replacement use case description and corresponding education levels}
\label{tab:use-case-description-long-labor}
\end{table*}
\begin{table*}[h!]
\footnotesize
\centering
\begin{tabularx}{\textwidth}{l|l|X}
\hline
\textbf{Use Case} & \textbf{EU Risk Level} & \textbf{Description} \\
\hline

Digital Medical Advice AI & High &
Imagine an AI system that provides preliminary medical assessments to help patients get efficient medical consultations and treatment (i.e., medical advice). It analyzes physical characteristics, reported symptoms, and medical history to suggest potential health issues, provide treatment plans, or direct patients to appropriate medical professionals and facilities. In emergencies, it can also act as a triage tool used by medical institutions to prioritize care. The system can read, write, and talk fluently. It has expert-level knowledge, can browse the internet, and can process and read massive amounts of data. We will refer to the AI system as Digital Medical Advice AI in the rest of the survey. \\
\hline

Customized Lifestyle Coach AI & High / Limited &
Imagine an AI system that offers personalized advice on how to manage healthy lifestyles and enhance wellness (i.e., customized lifestyle coaching). Beyond health tracking, it provides actionable insights on integrating fitness routines, dietary adjustments, and mindfulness practices into your daily schedule. The health data it collects can also be used to provide accurate information for life and health insurance. The system can read, write, and talk fluently. It has expert-level knowledge, can browse the internet, and can process and read massive amounts of data. We will refer to the AI system as Customized Lifestyle Coach AI in the rest of the survey. \\
\hline

Personal Health Research AI & Limited &
Imagine an AI system that summarizes and documents complex research findings related to personal health issues users are interested in (i.e., personalized health research findings). It translates medical jargon into accessible language, explains the latest studies relevant to conditions of user interest, and provides a summary of the overall findings. The system can read, write, and talk fluently. It has expert-level knowledge, can browse the internet, and can process and read massive amounts of data. We will refer to the AI system as Personal Health Research AI in the rest of the survey. \\
\hline

Nutrition Optimizer AI & Limited / Low &
Imagine an AI system that organizes meal schedules and optimizes nutritional intake based on a user’s specific health goals (i.e., optimal nutrition plan). It crafts recipes, meal plans, and grocery lists optimized for users’ culinary preferences, dietary needs, fitness ambitions, and lifestyle. The system can read, write, and talk fluently. It has expert-level knowledge, can browse the internet, and can process and read massive amounts of data. We will refer to the AI system as Nutrition Optimizer AI in the rest of the survey. \\
\hline

Flavorful Swaps AI & Low &
Imagine an AI system that aims to enhance users’ diets by suggesting delicious, health-conscious alternatives to unhealthy favorite foods (i.e., healthy flavorful swaps). It offers meal swaps that maintain satisfying flavors while being aligned with users’ health and dietary goals. The system can read, write, and talk fluently. It has expert-level knowledge, can browse the internet, and can process and read massive amounts of data. We will refer to the AI system as Flavorful Swaps AI in the rest of the survey. \\
\hline

\end{tabularx}
\caption{Personal health use cases and corresponding EU risk levels}
\label{tab:use-case-descriptions-long-personal-health}
\end{table*}

\subsection{Participant Details}
\label{app:participant-details}
The demographics of the participants for our study is shown in Tables \ref{app:demographics-1-jobs-p1} to \ref{app:demographics-3-personal-p2}. There was a fairly balanced distribution of participants across the different age groups, although there was a slightly higher proportion of participants in the 25-34 years old and 45-54 years old age ranges. In terms of racial distribution, there were more White/Caucasian participants compared to the other races. The gender distribution was relatively balanced in terms of males vs non-males. The participants were mostly employed or looking for work and a majority of them also had at least some form of college education. Most participants identified as liberal in terms of political leaning. Participants' AI literacy scores are shown in Table~\ref{tab:ai-literacy-distr} and AI Ethics score are shown in Table~\ref{tab:ai-ethics-distr}.

Participants were allocated 5 use cases from one of the scenarios and the allocation between the 2 scenarios are well-balanced and can be found in Table \ref{tab:study1-use-case}.
\begin{table}[!hbpt]
    \centering
    \small
    \begin{tabularx}{\linewidth}{l|X}
    \toprule
       Use Case & Participants Allocated \\
    \midrule
       Personal Use Cases & \\
    \midrule
       Digital Medical Advice & \multirow{5}{*}{97} \\
       Customized Lifestyle Coach & \\
       Personal Health Research & \\
       Nutrition Optimizer & \\
       Flavorful Swaps & \\
    \midrule
       Labor Replacement Use Cases & \\
    \midrule
       Lawyer & \multirow{5}{*}{100} \\
       Elementary School Teacher & \\
       IT Support Specialist & \\
       Government Eligibility Interviewer & \\
       Telemarketer & \\
    \bottomrule
    \end{tabularx}
    \caption{Participant allocation to each category of scenarios.}
    \label{tab:study1-use-case}
\end{table}

\begin{table*}[!htpb]
    \footnotesize
    \centering
    \begin{tabular}{{ll|ll|ll|ll}}
    \toprule
         \textbf{Racial Identity} & \textbf{\textit(N) (\%)} & \textbf{Age} & \textbf{\textit{N} (\%)} & \textbf{Gender Identity} & \textbf{\textit{N} (\%)} & \textbf{Education} & \textbf{\textit{N} (\%)} \\
         \midrule
        White or Caucasian & 33 (33.0) & 18-24 & 11 (11.0) & Man & 49 (49.0) & Bachelor’s degree & 36 (36.0)\\
        Black or African American & 23 (23.0) & 45-54 & 27 (27.0) & Non-male & 51 (51.0) & Graduate degree$^{*}$ & 18 (18.0)\\
        Asian & 21 (21.0) & 25-34 & 22 (22.0) & & & Some college $^{*}$ & 17 (17.0)\\
        Mixed & 13 (13.0) & 55-64 & 20 (20.0) & & & High school diploma$^{*}$ & 16 (16.0)\\
        Other & 10 (10.0) & 35-44 & 14 (14.0) & & & Associates degree$^{*}$ & 13 (13.0)\\
         & 2 (0.7) & 65+ & 6 (6.0) & & & Some high school$^{*}$ & 0 (0.0)\\
    \bottomrule
    \end{tabular}
    \caption{Labor Replacement Study 1 Survey: Racial, age, gender identities and education level of participants. Asterisk (*) denotes labels shortened due to space.}
    \label{app:demographics-1-jobs-p1}
\end{table*}
\begin{table*}[htpb]
    \centering
    \footnotesize
    \begin{tabular}{ll|ll|ll|ll}
    \toprule
         \textbf{Minority/Disadvantaged Group} & \textbf{\textit(N) (\%)} & \textbf{Transgender} & \textbf{\textit{N} (\%)} & \textbf{Sexuality} & \textbf{\textit{N} (\%)} & \textbf{Political Leaning} & \textbf{\textit{N} (\%)} \\
         \midrule
No & 68 (68.0) & No & 97 (97.0) & Heterosexual & 78 (78.0) & Liberal & 34 (34.0)\\
Yes & 32 (32.0) & Yes & 2 (2.0) & Others & 22 (22.0) & Moderate & 23 (23.0)\\
 &  & Prefer not to say & 1 (1.0) & & & Strongly liberal & 20 (20.0)\\
 &  &  &  &  &  & Conservative & 18 (18.0)\\
 &  &  &  &  &  & Strongly conservative & 4 (4.0)\\
 &  &  &  &  &  & Prefer not to say & 1 (1.0)\\
\bottomrule
    \end{tabular}
    \caption{Labor Replacement Study 1 Survey: Additional demographic identities}
    \label{app:demographics-2-jobs-p1}
\end{table*}
\begin{table*}[htpb]
    \centering
    \footnotesize
    \begin{tabularx}{\textwidth}{Xl|Xl|Xl|Xl}
    \toprule
    \textbf{Longest Residence} & \textbf{\textit(N) (\%)} & \textbf{Employment} & \textbf{\textit{N} (\%)} & \textbf{Occupation (Top 10)} & \textbf{\textit{N} (\%)} & \textbf{Religion} & \textbf{\textit{N} (\%)} \\
    \midrule
United States of America & 97 (97.0) & Employed, 40+ & 53 (53.0) & Other & 35 (35.0) & Christian & 29 (29.0)\\
Others & 3 (3.0) & Employed, 1-39 & 16 (16.0) & Prefer not to answer & 10 (10.0) & Agnostic & 20 (20.0)\\
 &  & Retired & 9 (9.0) & Health Care and Social Assistance & 10 (10.0) & Atheist & 15 (15.0)\\
 &  & Not employed, looking for work & 7 (7.0) & Information & 10 (10.0) & Nothing in particular & 13 (13.0)\\
 &  & Disabled, not able to work & 5 (5.0) & Manufacturing & 7 (7.0) & Catholic & 11 (11.0)\\
 &  & Not employed, NOT looking for work & 4 (4.0) & Professional, Scientific, and Technical Services & 7 (7.0) & Muslim & 5 (5.0)\\
 &  & Other: please specify & 4 (4.0) & Arts, Entertainment, and Recreation & 6 (6.0) & Hindu & 3 (3.0)\\
 &  & Prefer not to disclose & 2 (2.0) & Retail Trade & 6 (6.0) & Something else, Specify & 2 (2.0)\\
 &  &  &  & Finance and Insurance & 5 (5.0) & Jewish & 1 (1.0)\\
 &  &  &  & Transportation and Warehousing, and Utilities & 4 (4.0) & Buddhist & 1 (1.0)\\
 \bottomrule
    \end{tabularx}
    \caption{Labor Replacement Study 1 Survey: Additional demographic identities. The Occupation category was capped at the top 10 for brevity, with the remaining occupations merged together with the Other: please specify option.}
    \label{app:demographics-3-jobs-p1}
\end{table*}
\begin{table*}[!htpb]
    \footnotesize
    \centering
    \begin{tabular}{{ll|ll|ll|ll}}
    \toprule
         \textbf{Racial Identity} & \textbf{\textit(N) (\%)} & \textbf{Age} & \textbf{\textit{N} (\%)} & \textbf{Gender Identity} & \textbf{\textit{N} (\%)} & \textbf{Education} & \textbf{\textit{N} (\%)} \\
         \midrule
        White or Caucasian & 29 (29.9) & 45-54 & 30 (30.9) & Man & 50 (51.5) & Bachelor’s degree & 40 (41.2)\\
        Black or African American & 26 (26.8) & 25-34 & 26 (26.5) & Non-male & 47 (48.5) & Some college $^{*}$ & 21 (21.6)\\
        Asian & 20 (20.6) & 55-64 & 14 (14.4) & & & Graduate degree$^{*}$ & 14 (14.4)\\
        Other & 14 (14.4) & 35-44 & 13 (13.4) & & & High school diploma$^{*}$ & 13 (13.4)\\
        Mixed & 8 (8.2) & 18-24 & 9 (9.3) & & & Associates degree$^{*}$ & 8 (8.2)\\
         & 2 (0.7) & 65+ & 4 (4.1) & & & Some high school$^{*}$ & 1 (1.0)\\
         &  & Prefer not to disclose & 1 (1.0) & & & & \\
    \bottomrule
    \end{tabular}
    \caption{Personal Use Cases Study 1 Survey: Racial, age, gender identities and education level of participants. Asterisk (*) denotes labels shortened due to space.}
    \label{app:demographics-1-personal-p1}
\end{table*}
\begin{table*}[htpb]
    \centering
    \footnotesize
    \begin{tabular}{ll|ll|ll|ll}
    \toprule
         \textbf{Minority/Disadvantaged Group} & \textbf{\textit(N) (\%)} & \textbf{Transgender} & \textbf{\textit{N} (\%)} & \textbf{Sexuality} & \textbf{\textit{N} (\%)} & \textbf{Political Leaning} & \textbf{\textit{N} (\%)} \\
         \midrule
No & 51 (52.6) & No & 94 (96.9) & Heterosexual & 75 (77.3) & Liberal & 34 (35.1)\\
Yes & 46 (47.4) & Yes & 3 (3.1) & Others & 22 (22.7) & Moderate & 31 (32.0)\\
 &  & Prefer not to say & 0 (0.0) & & & Strongly liberal & 12 (12.4)\\
 &  &  &  &  &  & Conservative & 10 (10.3)\\
 &  &  &  &  &  & Strongly conservative & 9 (9.3)\\
 &  &  &  &  &  & Prefer not to say & 1 (1.0)\\
\bottomrule
    \end{tabular}
    \caption{Personal Use Cases Study 1 Survey: Additional demographic identities}
    \label{app:demographics-2-personal-p1}
\end{table*}
\begin{table*}[htpb]
    \centering
    \footnotesize
    \begin{tabularx}{\textwidth}{Xl|Xl|Xl|Xl}
    \toprule
        \textbf{Longest Residence} & \textbf{\textit(N) (\%)} & \textbf{Employment} & \textbf{\textit{N} (\%)} & \textbf{Occupation (Top 10)} & \textbf{\textit{N} (\%)} & \textbf{Religion} & \textbf{\textit{N} (\%)} \\
    \midrule
United States of America & 93 (95.9) & Employed, 40+ & 46 (47.4) & Other & 36 (35.6) & Christian & 40 (40.8)\\
Others & 4 (4.1) & Employed, 1-39 & 22 (22.7) & Health Care and Social Assistance & 11 (11.3) & Catholic & 16 (16.3)\\
 &  & Not employed, looking for work & 13 (13.4) & Prefer not to answer & 10 (10.3) & Agnostic & 15 (15.3)\\
 &  & Not employed, NOT looking for work & 4 (4.1) & Professional, Scientific, and Technical Services & 9 (9.3) & Nothing in particular & 11 (11.2)\\
 &  & Disabled, not able to work & 4 (4.1) & Educational Services & 9 (9.3) & Atheist & 5 (5.1)\\
 &  & Other: please specify & 4 (4.1) & Finance and Insurance & 8 (8.2) & Something else, Specify & 5 (5.1)\\
 &  & Retired & 3 (3.1) & Arts, Entertainment, and Recreation & 5 (5.2) & Buddhist & 3 (3.1)\\
 &  & Prefer not to disclose & 1 (1.0) & Manufacturing & 5 (5.2) & Muslim & 1 (1.0)\\
 &  &  &  & Retail Trade & 4 (4.1) & Jewish & 1 (1.0)\\
 &  &  &  & Accommodation and Food Services & 4 (4.1) & Hindu & 1 (1.0)\\
 \bottomrule
    \end{tabularx}
    \caption{Personal Use Cases Study 1 Survey: Additional demographic identities. The Occupation category was capped at the top 10 for brevity, with the remaining occupations merged together with the Other: please specify option.}
    \label{app:demographics-3-personal-p1}
\end{table*}
\begin{table*}[!hbpt]
\centering
\small
\begin{tabular}{ccccccc}
\hline
Score & AI Awareness & AI Usage & AI Evaluation & Gen AI Usage Freq. & Gen AI Limit. Familiarity \\ \hline
1  & 25           & 15       & 30            & 35    & 55   \\
2  & 40           & 60       & 75            & 200   & 320  \\
3  & 75           & 90       & 105           & 155   & 345  \\
4  & 140          & 125      & 125           & 235   & 230  \\
5  & 380          & 310      & 275           & 220   & 35   \\
6  & 280          & 300      & 310           & 140   &  \NA    \\
7  & 45           & 85       & 65            &     \NA &    \NA \\
\hline
\end{tabular}
\caption{AI literacy scale participant count. Questions are on a 7-point likert scale of increasing score meaning increase in literacy for the aspect. Gen AI Usage Frequency has max score of 6 and Limitation Familiarity has max value of 5.}
\label{tab:ai-literacy-distr}
\end{table*}
\begin{table}[h]
\centering
\small
\begin{tabular}{lc}
\hline
AI Ethics Score & Count \\ \hline
5  & 15             \\
6  & 10             \\
7  & 25             \\
8  & 25             \\
9  & 75             \\
10 & 105            \\
11 & 165            \\
12 & 100            \\
13 & 105            \\
14 & 90             \\
15 & 120            \\
16 & 80             \\
17 & 35             \\
18 & 35             \\ \hline
\end{tabular}
\caption{AI ethics score count for total AI ethics score (sum over 3 questions with 7 point likert scale with max possible value of 21)}
\label{tab:ai-ethics-distr}
\end{table}
\subsection{Open-text Annotation Dimensions}
\label{sssec:reasoning-dim}
\paragraph{Reasoning Type}
Inspired by previous works in moral psychology, we used two main reasoning types to characterize participants' decision making pattern as expressed in their open-text answers: cost-benefit reasoning and rule-based reasoning \citep{cheung2024measuring}. These two reasoning types are rooted in two decision making processes in moral and wider decision making literature: utilitarian and deontological reasoning, respectively. Cost-benefit reasoning thus considers the possible outcomes and their expected utility or value when making decisions, and rule-based reasoning shows more inherent value in action or entities. See \S~\ref{ssec:moral-decision-making} for further discussion.

\paragraph{Moral Values}
To annotate which values were prevalent in participants' considerations of use cases, we used five moral values: care, fairness, loyalty, authority, and purity \citep{graham2011mapping,graham2008moral}. While these dimensions have been re-defined to include more diverse values from participants beyond WEIRD (white, educated, industrialized, rich, and democratic) \citep{atari2023morality}, we used these five dimensions due to brevity of the questionnaire, which was used in the survey to provide importance of each values to participants. 

\paragraph{Switching Conditions}
We annotated concerns expressed in switching conditions using three categories: functionality (e.g., errors, bias in systems, limited capabilities), usage (e.g., errors, bias in systems, limited capabilities), and societal impact (e.g., job loss, over-reliance), inspired by harm taxonomy developed by \citeauthor{solaiman2023evaluating} and user concern annotation practice adopted by \citeauthor{mun2024participaidemocraticsurveyingframework}.

\section{Extended Results}
\label{app:extended_results}
\begin{figure*}[!htbp]
    \centering
    \includegraphics[width=\textwidth]{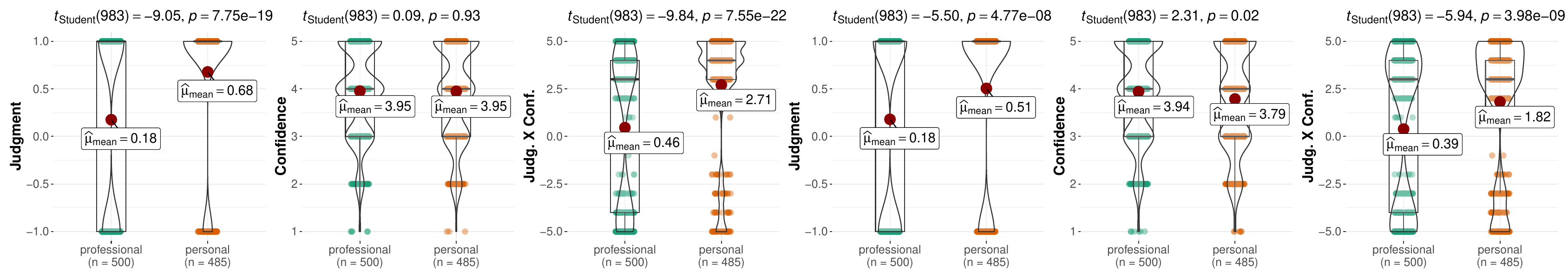}
    \caption{Use case category means and distributions of numerically converted Judgment, Confidence, and Judgment$\times$Confidence. Left three show results for development decisions (Q1, Q2) and right three show results for usage decisions (Q5, Q6). Significance was calculated using Student's t-test as indicated by labels above each plot.
    }
    \label{fig:use-case-effect-category}
\end{figure*}

Figure~\ref{fig:use-case-effect-category} shows the distribution of participants' judgment variation across categories as discussed in \S\ref{sec:RQ1}.

\section{Open-text Annotation Details}
\label{app:open-text-annotation-details}
\subsection{Automatic Annotation}
\subsubsection{Methods}
We used Open-AI's gpt-4o model with maximum tokens set to 1024 to control response length, use a temperature of 0.7 to manage randomness, and keep top\_p at 1 with default settings for frequency and presence penalties at 0. Prompts will be released with data upon acceptance. 

\subsubsection{Results}
Results for inter-rater reliability between human and gpt-4o annotations are shown in Table~\ref{tab:irr-results}.
\begin{table*}[!hptb]
\centering
\begin{tabular}{l|cccccccc}
\hline
\textbf{Category} & \textbf{AC1} & \textbf{Interpretation} & \textbf{95\% CI} & \textbf{p-value} & \textbf{z} & \textbf{SE} & \textbf{PA} & \textbf{PE} \\ \hline
Cost Benefit & 0.976 & Almost Perfect & (0.942, 1.000) & \textbf{0.0} & 56.9 & 0.0172 & 0.980 & 0.164 \\ 
Rule Based & 0.848 & Almost Perfect & (0.754, 0.943) & \textbf{0.0} & 17.8 & 0.0476 & 0.890 & 0.276 \\
Care             & 0.705 & Substantial        & (0.563, 0.847)     & $\mathbf{2.22 \times 10^{-16}}$ & 9.88   & 0.0713  & 0.850 & 0.492 \\
Fairness         & 0.609 & Substantial        & (0.448, 0.769)     & $\mathbf{2.36 \times 10^{-11}}$ & 7.53   & 0.0808  & 0.790 & 0.464 \\
Authority        & 0.897 & Almost Perfect     & (0.822, 0.972)     & \textbf{0.0}           & 23.70  & 0.0378  & 0.920 & 0.226 \\
Purity           & 0.801 & Almost Perfect     & (0.693, 0.908)     & \textbf{0.0}           & 14.77  & 0.0542  & 0.850 & 0.248 \\
Functionality    & 0.701 & Substantial        & (0.559, 0.844)     & $\mathbf{4.44 \times 10^{-16}}$ & 9.79   & 0.0716  & 0.850 & 0.498 \\
Usage            & 0.674 & Substantial        & (0.527, 0.822)     & $\mathbf{1.27 \times 10^{-14}}$ & 9.05   & 0.0745  & 0.830 & 0.478 \\
Societal Impact  & 0.566 & Moderate           & (0.397, 0.735)     & $\mathbf{1.59 \times 10^{-9}}$  & 6.66   & 0.0851  & 0.750 & 0.424 \\
\hline
\end{tabular}
\caption{Inter-rater Agreement using Gwet's AC1. Interpretation according to \citep{wongpakaran2013comparison}.}
\label{tab:irr-results}
\end{table*}

\section{Factors Impacting Acceptability Judgments}
\subsection{Use Case Factors}
\begin{figure}[!hptb]
    \centering
    \includegraphics[width=\linewidth]{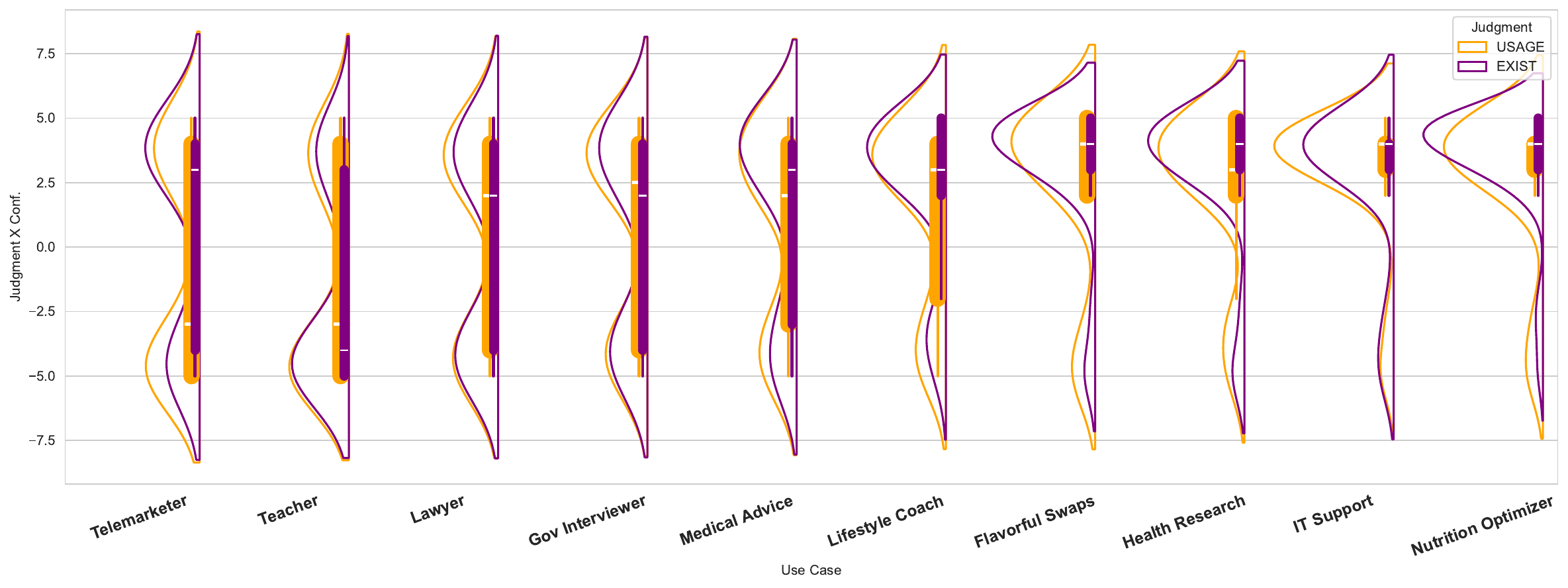}
    \caption{Numerically converted Judgment x Confidence (-5, 5) by use cases distributions sorted by standard deviation of both existence and usage (sum; highest to lowest) using data from Study 1 results.}
    \label{fig:decisions-by-use-case-violin}
\end{figure}

Additional analysis of use case factors showing distribution of judgments by use case sorted by standard deviation is shown in Figure~\ref{fig:decisions-by-use-case-violin}. Table~\ref{table:use-case-effect-anova} shows analysis of use case effect using ANOVA.
\begin{table*}[!hptb]
    \begin{center}
    \small
    \begin{tabular}{@{}llcccccc@{}}
        \toprule
        \textbf{Acceptability} & \textbf{Aspect} & \textbf{Factor} & \textbf{Sum Sq} & \textbf{Mean Sq} & \textbf{NumDF} & \textbf{DenDF} & \textbf{Pr($>$F)} \\ 
        \midrule
        \multirow{4}{*}{EXIST} & \multirow{2}{*}{Judgment} & Category & 29.903 & 29.903 & 1 & 197 & \textbf{5.98e-11} ***\\
        & & Use Case & 86.116 & 9.5684 & 9 & 641.38 & \textbf{$<$ 2.2e-16} ***\\ \cmidrule{2-8}
        & \multirow{2}{*}{Confidence} & Category & 0.0017113 & 0.0017113 & 1 & 197 & 0.9563 \\
        & & Use Case & 13.519 & 1.5021 & 9 & 603.23  2.7243 & \textbf{0.004037} **\\ \midrule
        \multirow{4}{*}{USAGE} & \multirow{2}{*}{Judgment} & Category & 8.4257 & 8.4257 & 1 & 197 & \textbf{0.0002488} ***\\
        & & Use Case & 73.801 & 8.2001 & 9 & 610.34 & \textbf{$<$ 2.2e-16} ***\\ \cmidrule{2-8}
        & \multirow{2}{*}{Confidence} & Category & 1.3444 & 1.3444 & 1 & 197 & 0.153 \\
        & & Use Case & 20.721 & 2.3023 & 9 & 603.36 & \textbf{0.0001783} ***\\
        \bottomrule
    \end{tabular}
    \caption{ANOVA analysis of LMER models \texttt{judgment $\sim$ category + (1 | subject)} and \texttt{judgment $\sim$ useCase + (1 | subject)} (same formulas repeated with \texttt{confidence} as a dependent variable){} analyzed with Study 1 data.}
    \label{table:use-case-effect-anova}
    \end{center}
\end{table*}

\subsection{Demographics Factors}
Additional analysis using ANOVA for demographic factors is shown in Table~\ref{tab:demographics-anova}.
\begin{table*}[!hptb]
\centering
\small
\begin{tabular}{lcccccc}
\hline
 & \multicolumn{3}{c}{EXIST} & \multicolumn{3}{c}{USAGE}\\
\cmidrule(lr){2-4}\cmidrule(lr){5-7}
\textbf{Demographics} & Judgment & Confidence & Judg.$\times$Conf. & Judgment & Confidence & Judg$\times$Conf.\\
\hline
\multicolumn{7}{l}{\textbf{All}}\\
\quad Gender & $\mathbf{16.60^{***}}$ & $0.19$ &  $\mathbf{16.71^{***}}$ & $\mathbf{15.26^{***}}$ & $0.83$ & $\mathbf{13.14^{***}}$ \\
\quad Race               & $1.62$ & $\mathbf{4.09^{**}}$ & $1.45$ & $0.65$ & $\mathbf{5.12^{***}}$ & $0.45$ \\
\quad Employment         & $1.13$ & $\mathbf{3.03^{*}}$ & $1.14$ & $0.43$ & $1.71$ & $0.69$ \\
\quad Sexual Orientation & $0.42$ & $0.37$ & $0.19$ & $0.75$ & $\mathbf{5.22^{*}}$ & $0.09$ \\
\hline
\multicolumn{7}{l}{\textbf{Professional}} \\
\quad Race               & $\mathbf{2.56^{*}}$   & $1.80$                & $\mathbf{2.34^{.}}$    & $1.04$                & $\mathbf{2.91^{*}}$   & $0.48$                \\
\quad Gender             & $\mathbf{18.37^{***}}$& $0.05$                & $\mathbf{19.51^{***}}$ & $\mathbf{19.83^{***}}$& $0.19$                & $\mathbf{20.21^{***}}$\\
\quad Education          & $1.98$                & $0.96$                & $1.34$                 & $2.25^{.}$            & $1.12$                & $2.07^{.}$            \\
\quad Discrimination & $2.18$                & $0.68$                & $2.67^{.}$             & $0.29$                & $1.46$                & $0.13$                \\
\hline
\multicolumn{7}{l}{\textbf{Personal}}\\
\quad Race               & $2.11^{.}$ & $\mathbf{4.36^{**}}$  & $2.28^{.}$  & $0.16$ & $\mathbf{4.07^{**}}$ & $0.38$ \\
\quad Political View     & $0.38$     & $\mathbf{3.39^{*}}$   & $0.86$      & $0.33$ & $1.56$ & $0.36$ \\
\quad Employment         & $0.85$     & $\mathbf{2.42^{*}}$   & $1.47$      & $0.33$ & $0.30$ & $0.36$ \\
\hline
\multicolumn{7}{l}{\scriptsize{$^{***}p<0.001$; $^{**}p<0.01$; $^{*}p<0.05$; $^{.}p<0.1$}}
\end{tabular}
\caption{ANOVA Results by Demographic Category (F‐value with Significance)}
\label{tab:demographics-anova}
\end{table*}

\begin{table*}[hpbt]
\begin{center}
\small
\begin{tabular}{l c c c c c c}
\hline
 & \multicolumn{3}{c}{EXIST ($\beta$ (SE))} & \multicolumn{3}{c}{USAGE ($\beta$ (SE))}\\
\cmidrule(lr){2-4}\cmidrule(lr){5-7}
Decision Style Factors & Judg. & Conf. & Judg.$\times$Conf. & Judg. & Conf. & Judg.$\times$Conf. \\
\hline
(Intercept) & $0.11 \ (0.34)$ & $\mathbf{3.10^{***}} \ (0.46)$ & $-0.32 \ (1.46)$ & $-0.74 \ (0.39)$ & $\mathbf{2.98^{***}} \ (0.49)$ & $\mathbf{-4.04^{*}} \ (1.68)$ \\
MFQ Care & $0.00 \ (0.01)$ & $-0.01 \ (0.02)$ & $0.01 \ (0.06)$ & $-0.00 \ (0.02)$ & $0.02 \ (0.02)$ & $-0.01 \ (0.07)$ \\
MFQ Fairness & $-0.01 \ (0.02)$ & $0.04 \ (0.02)$ & $-0.03 \ (0.07)$ & $0.01 \ (0.02)$ & $0.01 \ (0.02)$ & $0.05 \ (0.08)$ \\
MFQ Loyalty & $\mathbf{0.04^{***}} \ (0.01)$ & $-0.01 \ (0.02)$ & $\mathbf{0.18^{***}} \ (0.05)$ & $\mathbf{0.04^{**}} \ (0.01)$ & $-0.01 \ (0.02)$ & $\mathbf{0.18^{**}} \ (0.06)$ \\
MFQ Authority & $-0.01 \ (0.01)$ & $0.02 \ (0.02)$ & $-0.05 \ (0.05)$ & $0.00 \ (0.02)$ & $0.03 \ (0.02)$ & $0.02 \ (0.06)$ \\
MFQ Purity & $0.00 \ (0.01)$ & $0.02 \ (0.01)$ & $0.04 \ (0.04)$ & $-0.01 \ (0.01)$ & $0.00 \ (0.02)$ & $-0.01 \ (0.05)$ \\
Empathy & $0.01 \ (0.03)$ & $0.03 \ (0.04)$ & $0.05 \ (0.13)$ & $0.06 \ (0.04)$ & $0.02 \ (0.05)$ & $0.28 \ (0.15)$ \\
InstrumentalHarm & $0.00 \ (0.03)$ & $-0.01 \ (0.04)$ & $0.04 \ (0.13)$ & $-0.02 \ (0.03)$ & $-0.01 \ (0.04)$ & $-0.06 \ (0.15)$ \\
ImpartialBenificence & $0.03 \ (0.03)$ & $-0.01 \ (0.04)$ & $0.08 \ (0.13)$ & $0.02 \ (0.04)$ & $-0.02 \ (0.05)$ & $0.06 \ (0.15)$ \\
\hline
AIC & $2412.17$ & $2539.14$ & $5149.55$ & $2442.29$ & $2671.82$ & $5132.26$ \\
BIC & $2470.88$ & $2597.85$ & $5208.26$ & $2501.01$ & $2730.53$ & $5190.97$ \\
Log Likelihood & $-1194.08$ & $-1257.57$ & $-2562.77$ & $-1209.15$ & $-1323.91$ & $-2554.13$ \\
Num. obs. & $985$ & $985$ & $985$ & $985$ & $985$ & $985$ \\
Num. groups: prolific\_id & $197$ & $197$ & $197$ & $197$ & $197$ & $197$ \\
Num. groups: use\_case & $10$ & $10$ & $10$ & $10$ & $10$ & $10$ \\
Var: prolific\_id (Intercept) & $0.12$ & $0.37$ & $2.71$ & $0.23$ & $0.42$ & $4.65$ \\
Var: use\_case (Intercept) & $0.11$ & $0.01$ & $2.22$ & $0.08$ & $0.02$ & $1.58$ \\
Var: Residual & $0.55$ & $0.56$ & $8.53$ & $0.53$ & $0.64$ & $7.70$ \\
\hline
\multicolumn{7}{l}{\scriptsize{$^{***}p<0.001$; $^{**}p<0.01$; $^{*}p<0.05$}}
\end{tabular}
\caption{Coefficients with standard error in parenthesis with following models: \texttt{Judgment $\sim$ $\texttt{MFQ}_{foundation}$ + Empathy + InustrumentalHarm + ImpartialBeneficence + (1|Subject) + (1|useCase)}. Bolded value for empathy had $p<0.1$.}
\label{tab:reasoning-factors-questionnaires}
\end{center}
\end{table*}
\subsubsection{Questionnaires}
Interestingly, only Loyalty had a significant effect on both existence ($0.20, p<.001$) and usage ($0.20, p<.01$) as shown in Table~\ref{tab:reasoning-factors-questionnaires}. Moreover, Empathy had a positive and marginally significant effect for usage ($.09, p<.1$). However, Loyalty, as shown in Figure~\ref{fig:moral_values_proportions}, does not appear as frequently in participants' open text responses compared to values such as Care and Fairness and is the only value that did not have a significant association with use cases. 

\section{Factors in Participant Rationale}

\subsection{Reasoning Types}
\begin{figure*}[!hptb]
    \centering
    \subfigure[Acceptance judgment and reasoning type mapping throughout professional use cases.]{
        \includegraphics[width=\linewidth]{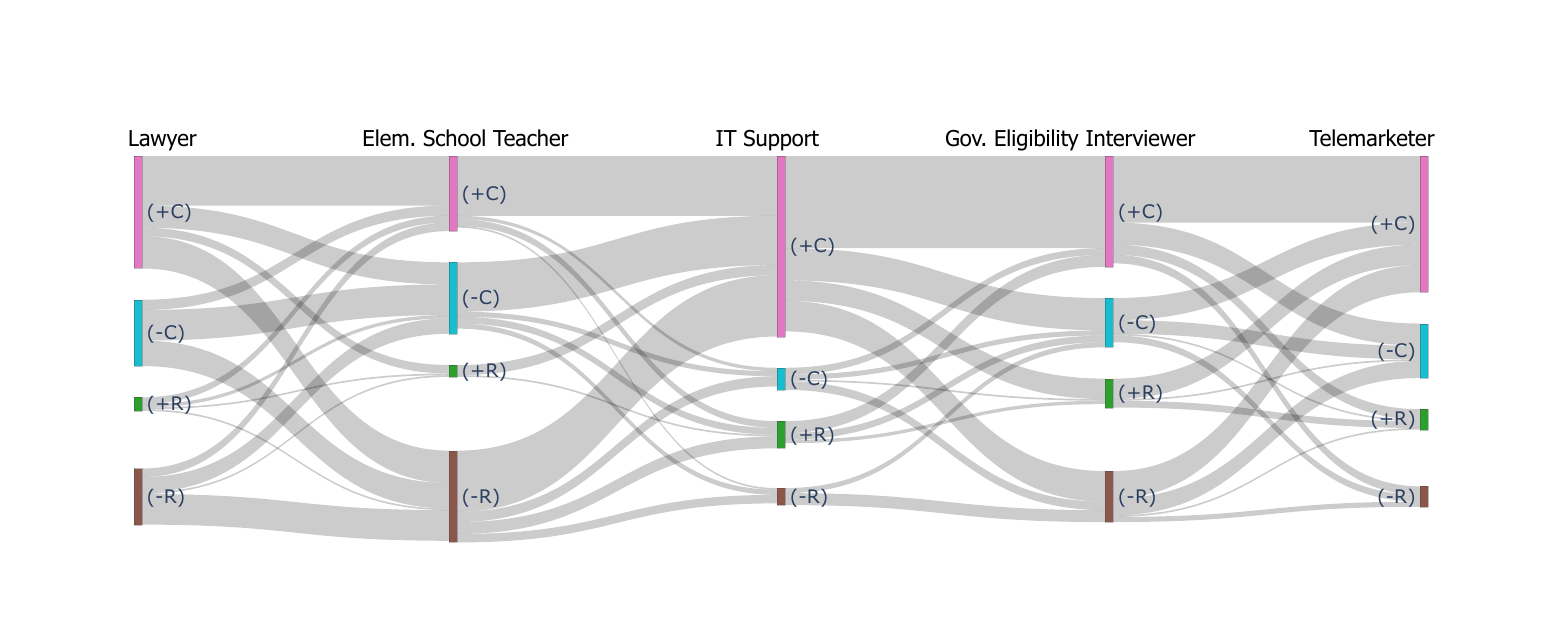}
        \label{fig:subfig1}
    }
    \hfill
    \subfigure[Acceptance judgment and reasoning type mapping throughout personal use cases.]{
        \includegraphics[width=\linewidth]{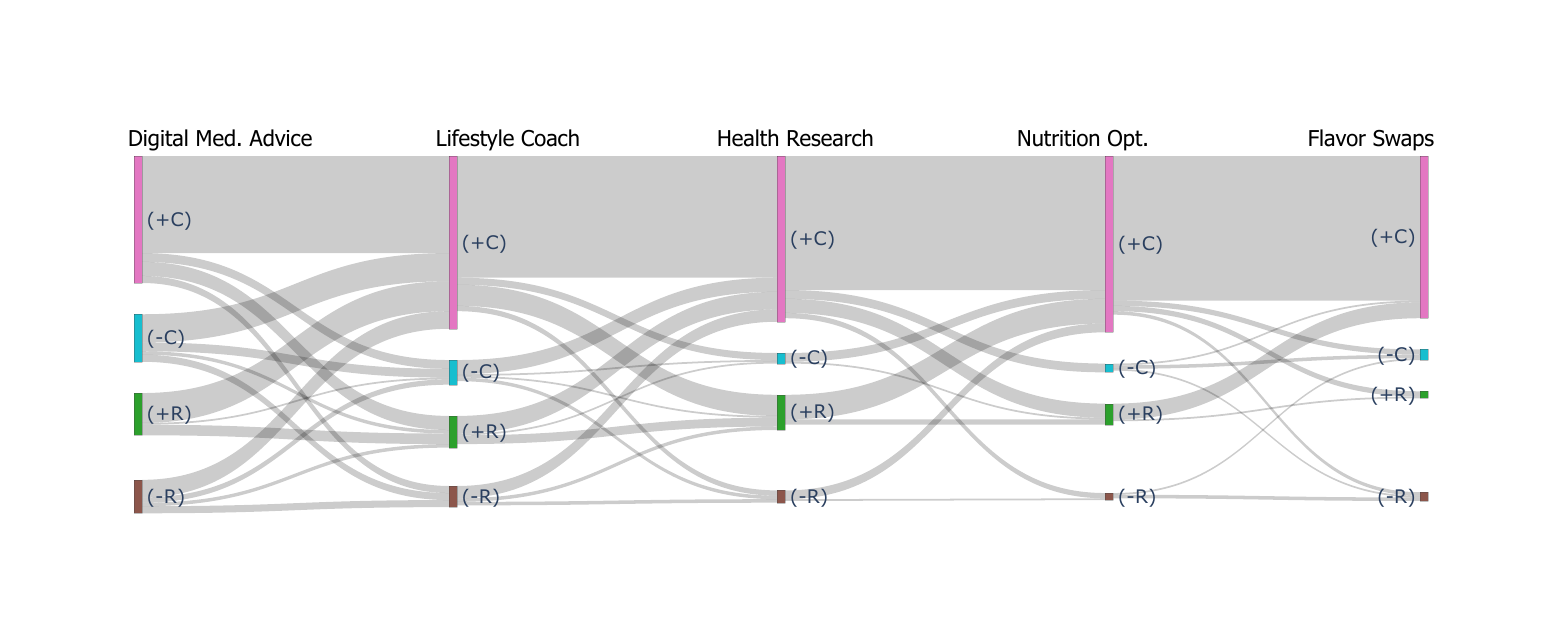}
        \label{fig:subfig2}
    }
    \caption{Mapping of decisions and reasoning types. + and - denote positive and negative acceptance. ``C'' denotes Cost-benefit analysis and ``R'' denotes Rule-based reasoning.}
    \label{fig:reasoning-type-sankey}
\end{figure*}

We show the flow of participants' decisions and corresponding rationales throughout use cases in Figure~\ref{fig:reasoning-type-sankey}, which shows interesting distribution and switching of reasoning types, which would be interesting for future studies to consider. Moreover, Table~\ref{tab:reasoning-type-questionnaire} shows that there are almost no relation between reasoning types used by the participants and the decision-making style questionnaire results signifying that the reasoning types might be highly use-case specific rather than a character trait. It would be interesting to study the factors that actually influence the choice of reasoning types. 
\begin{table}
\begin{center}
\begin{tabular}{l c c}
\hline
 & Cost-benefit & Rule-based \\
\hline
(Intercept)                   & $\mathbf{0.73^{***}}(0.13)$ & $\mathbf{0.28^{*}}(0.13)$ \\
MFQ Care                     & $-0.00(0.01)$      & $0.00(0.01)$   \\
MFQ Fairness                 & $0.01(0.01)$       & $0.00(0.01)$   \\
MFQ Loyalty                  & $\mathbf{0.01^{*}}(0.00)$   & $-0.01(0.01)$ \\
MFQ Authority                & $-0.00(0.01)$      & $0.00(0.01)$   \\
MFQ Purity                   & $-0.00(0.00)$      & $0.00(0.00)$   \\
Empathy                & $0.00(0.01)$       & $-0.01(0.01)$  \\
InstrumentalHarm              & $0.00(0.01)$       & $-0.00(0.01)$  \\
ImpartialBenificence          & $-0.00(0.01)$      & $-0.00(0.01)$  \\
\hline
AIC                           & $668.43$     & $815.32$   \\
BIC                           & $727.14$     & $874.03$   \\
Log Likelihood                & $-322.21$    & $-395.66$  \\
Num. obs.                     & $985$        & $985$      \\
Num. groups: prolific\_id     & $197$        & $197$      \\
Num. groups: use\_case        & $10$         & $10$       \\
Var: prolific\_id (Intercept) & $0.02$       & $0.02$     \\
Var: use\_case (Intercept)    & $0.00$       & $0.01$     \\
Var: Residual                 & $0.10$       & $0.11$     \\
\hline
\multicolumn{3}{l}{\scriptsize{$^{***}p<0.001$; $^{**}p<0.01$; $^{*}p<0.05$}}
\end{tabular}
\caption{Coefficient and standard error with significance. Model defined by \texttt{reasoningType $\sim$ MFQ$_{foundation}$ + Empathy + InstrumentalHarm + ImpartialBeneficence + (1|subject) + (1|useCase)}}
\label{tab:reasoning-type-questionnaire}
\end{center}
\end{table}

\subsection{Impact of Rationale Factors on Judgment}
We display the analysis results using ANOVA to understand the effect of rationale factors on judgment in Table~\ref{tab:reasoning-factors-judgment-anova}.
\begin{table*}[!hptb]
    \begin{center}
    \small
    \begin{tabular}{@{}llccccc@{}}
        \toprule
        \textbf{Acceptability} & \textbf{Factor} & \textbf{Sum Sq} & \textbf{Mean Sq} & \textbf{NumDF} & \textbf{DenDF} & \textbf{Pr($>$F)} \\ 
        \midrule
        \multirow{16}{*}{\textbf{EXIST}} 
        & \multicolumn{5}{l}{\textbf{Judgment}} \\ 
        & \quad Cost Benefit & 2.764 & 2.764 & 1 & 953.92 & \textbf{0.0017} ** \\ 
        & \quad Rule Based & 6.939 & 6.939 & 1 & 937.59 & \textbf{7.57e-07} *** \\ 
        & \quad Fairness & 3.073 & 3.073 & 1 & 969.63 & \textbf{0.00095} *** \\ 
        & \quad Authority & 3.941 & 3.941 & 1 & 973.72 & \textbf{0.00018} *** \\ 
        & \quad Usage & 127.631 & 127.631 & 1 & 828.54 & \textbf{$<$2.2e-16} *** \\ 
        \addlinespace
        & \multicolumn{5}{l}{\textbf{Confidence}} \\
        & \quad Rule Based & 5.5845 & 5.5845 & 1 & 854.68 & \textbf{0.00149} ** \\ 
        & \quad Fairness & 3.9022 & 3.9022 & 1 & 893.72 & \textbf{0.00787} ** \\ 
        \addlinespace
        & \multicolumn{5}{l}{\textbf{Judgment x Confidence}} \\
        & \quad Cost Benefit & 44.72 & 44.72 & 1 & 936.85 & \textbf{0.00072} *** \\ 
        & \quad Rule Based & 82.21 & 82.21 & 1 & 917.67 & \textbf{4.79e-06} *** \\
        & \quad Fairness & 43.81 & 43.81 & 1 & 971.97 & \textbf{0.00081} *** \\ 
        & \quad Authority & 56.43 & 56.43 & 1 & 965.79 & \textbf{0.00015} *** \\ 
        & \quad Usage & 2458.62 & 2458.62 & 1 & 887.02 & \textbf{$<$2.2e-16} *** \\ 
        & \quad Societal Impact & 22.15 & 22.15 & 1 & 969.86 & \textbf{0.0171} *  \\ 
        \midrule
        \multirow{12}{*}{\textbf{USAGE}} 
        & \multicolumn{5}{l}{\textbf{Judgment}} \\ 
        & \quad Cost Benefit & 0.28 & 0.28 & 1 & 930.56 & \textbf{0.0177} * \\ 
        & \quad Usage & 496.98 & 496.98 & 1 & 937.06 & \textbf{$<$2e-16} *** \\ 
        \addlinespace
        & \multicolumn{5}{l}{\textbf{Confidence}} \\ 
        & \quad Cost Benefit & 2.4825 & 2.4825 & 1 & 863.46 & \textbf{0.0443} * \\ 
        & \quad Care & 3.6966 & 3.6966 & 1 & 812.12 & \textbf{0.0142} * \\ 
        & \quad Usage & 11.9628 & 11.9628 & 1 & 813.76 & \textbf{1.12e-05} *** \\ 
        \addlinespace
        & \multicolumn{5}{l}{\textbf{Judgment x Confidence}} \\ 
        & \quad Cost Benefit & 143.66 & 143.66 & 1 & 872.38 & \textbf{4.92e-06} *** \\ 
        & \quad Fairness & 72.485 & 72.485 & 1 & 918.88 & \textbf{0.00113} ** \\ 
        & \quad Societal Impact & 69.555 & 69.555 & 1 & 932.22 & \textbf{0.00143} ** \\ 
        \bottomrule
    \end{tabular}
    \caption{ANOVA analysis of the LMER model results (significant results only).}
    \label{tab:reasoning-factors-judgment-anova}
    \end{center}
\end{table*}

\subsection{Factors Influencing Moral Foundations in  Rationale}
In Table~\ref{tab:reasoning-fators-moral-values-annotations} we display analysis result using linear mixed effects on factors that may influence moral foundations appealed to in participants' rationales. We find greater relations with the use cases than personal factors. 
\begin{table*}
\begin{center}
\small
\begin{tabular}{l c c c c c}
\hline
 & Care & Fairness & Purity & Loyalty & Authority \\
\hline
(Intercept) (Telemarketer)                                 & $\mathbf{1.74^{***} \;(0.31)}$ & $0.29 \;(0.23)$       & $\mathbf{-2.80^{***} \;(0.42)}$ & $\mathbf{-4.82^{***} \;(1.03)}$  & $\mathbf{-2.29^{***} \;(0.35)}$ \\
MFQ\_care                                   & $0.08 \;(0.19)$       & $-0.12 \;(0.13)$      & $-0.36 \;(0.19)$       & $-0.15 \;(0.43)$        & $-0.01 \;(0.19)$       \\
MFQ\_fairness                               & $0.13 \;(0.19)$       & $\mathbf{0.31^{*} \;(0.13)}$   & $0.30 \;(0.20)$        & $-0.19 \;(0.43)$        & $0.14 \;(0.19)$        \\
MFQ\_loyalty                                & $\mathbf{0.60^{**} \;(0.21)}$  & $\mathbf{0.31^{*} \;(0.14)}$   & $-0.14 \;(0.21)$       & $-0.34 \;(0.57)$        & $-0.12 \;(0.20)$       \\
MFQ\_authority                              & $\mathbf{-0.48^{*} \;(0.24)}$  & $-0.23 \;(0.16)$      & $0.23 \;(0.24)$        & $0.06 \;(0.56)$         & $0.23 \;(0.24)$        \\
MFQ\_purity                                 & $0.28 \;(0.19)$       & $-0.16 \;(0.13)$      & $-0.07 \;(0.20)$       & $-0.05 \;(0.44)$        & $-0.22 \;(0.19)$       \\
empathy\_total                              & $0.06 \;(0.15)$       & $0.04 \;(0.10)$       & $0.07 \;(0.15)$        & $0.24 \;(0.39)$         & $-0.05 \;(0.15)$       \\
InstrumentalHarm                            & $-0.04 \;(0.15)$      & $0.12 \;(0.10)$       & $-0.14 \;(0.16)$       & $-0.10 \;(0.39)$        & $0.09 \;(0.15)$        \\
ImpartialBenificence                        & $0.09 \;(0.15)$       & $-0.05 \;(0.10)$      & $\mathbf{-0.37^{*} \;(0.16)}$   & $-0.44 \;(0.41)$        & $-0.07 \;(0.15)$       \\
Gov. Eligi. Interviewer & $0.06 \;(0.39)$       & $\mathbf{1.45^{***} \;(0.35)}$ & $\mathbf{-1.72^{*} \;(0.82)}$   & \NA  & $-0.01 \;(0.44)$       \\
IT Support Specialist              & $\mathbf{1.14^{*} \;(0.46)}$   & $\mathbf{0.76^{*} \;(0.32)}$   & $0.44 \;(0.49)$        & \NA & $-0.72 \;(0.50)$       \\
Elementary School Teacher          & $\mathbf{0.88^{*} \;(0.44)}$   & $\mathbf{-0.66^{*} \;(0.31)}$  & $0.90 \;(0.47)$        & $1.43 \;(1.13)$         & $0.42 \;(0.42)$        \\
Lawyer                             & $-0.48 \;(0.37)$      & $\mathbf{0.70^{*} \;(0.32)}$   & $-0.71 \;(0.60)$       & $0.71 \;(1.24)$         & $\mathbf{1.41^{***} \;(0.40)}$  \\
Flavorful Swaps                    & $0.73 \;(0.47)$       & $-0.18 \;(0.33)$      & $\mathbf{1.50^{**} \;(0.48)}$   & $-0.04 \;(1.43)$        & $-0.80 \;(0.55)$       \\
Nutrition Optimizer                & $\mathbf{1.14^{*} \;(0.50)}$   & $-0.37 \;(0.33)$      & $-0.01 \;(0.55)$       & \NA & $-0.26 \;(0.50)$       \\
Personal Health Research           & $\mathbf{1.46^{**} \;(0.54)}$  & $0.53 \;(0.34)$       & $-0.32 \;(0.58)$       & \NA   & $0.42 \;(0.47)$        \\
Cust, Lifestyle Coach         & $0.71 \;(0.47)$       & $0.47 \;(0.34)$       & $0.56 \;(0.51)$        & $-0.04 \;(1.43)$        & $-0.64 \;(0.54)$       \\
Digital Medical Advice             & $\mathbf{1.45^{**} \;(0.53)}$  & $0.02 \;(0.33)$       & $-0.50 \;(0.60)$       & \NA & $\mathbf{1.47^{***} \;(0.44)}$  \\
\hline
AIC                                         & $750.75$     & $1256.07$    & $644.21$      & $120.53$       & $845.93$      \\
BIC                                         & $843.71$     & $1349.03$    & $737.17$      & $213.49$       & $938.89$      \\
Log Likelihood                              & $-356.38$    & $-609.04$    & $-303.11$     & $-41.26$       & $-403.97$     \\
Num. obs.                                   & $985$        & $985$        & $985$         & $985$          & $985$         \\
Num. groups: prolific\_id                   & $197$        & $197$        & $197$         & $197$          & $197$         \\
Var: prolific\_id (Intercept)               & $1.27$       & $0.63$       & $1.10$        & $0.00$         & $1.56$        \\
\hline
\multicolumn{6}{l}{\scriptsize{$^{***}p<0.001$; $^{**}p<0.01$; $^{*}p<0.05$}}
\end{tabular}
\caption{Effects and standard error in parenthesis of the annotation output of participant answers modeled with following formula \texttt{Annot$_{foundation}$ $\sim$ MFQ$_{foundation}$ + empathy + instrumentalHarm + impartialBeneficence + useCase + (1|subject)} using \texttt{glmer} with family set to binomial. Intercept shows effects when categorical variables are set to following: \texttt{useCase = Telemarketer} and \texttt{Type = Cost-Benefit}.}
\label{tab:reasoning-fators-moral-values-annotations}
\end{center}
\end{table*}

\section{Survey 2: Explicit Weighing of Harms and Benefits of Use Cases}
Although not included in main text, we administered a variation of our main study where we asked participants to explicitly reason through harms and benefits. The decisions were measured before and after the explicit weighing of harms and benefits. However, we saw almost no effect. 

\subsection{Study Overview}
To better understand the reasoning behind participants decisions about the judgment and usage of the use cases, we conducted a second study with 1 survey for each category (Labor Replacement Use Cases and Personal Use Cases). The second study includes an additional set of questions to elicit the harms and benefits of developing and not developing an use case to better understand the reasoning behind participants decisions. Furthermore, we asked participants the judgment and usage decision questions before and after the set of harms and benefits questions to see if listing out reasons about an use case would elicit any change in their decisions. The details of the second study can be found in \ref{sssec:a-details} and the results can be found in \ref{sssec:a-results}.

\subsection{Setup and Details}\label{sssec:a-details}
While these same set of questions are asked for all five use cases for our main study, in our second study, participants are randomly allocated a single use case. The second study differs from the main study with an initial set of judgment questions without open-text rationales (Q1 - Initial to Q4 - Initial), which are followed by explicit listing and weighing of the possible harms and benefits of the use case in the context of both developing and not developing the use case. We then again ask participants the same set of judgment questions along with the open-text questions to elaborate on their reasoning, similar to the main study. To understand how the judgment and usage decisions are affected by other factors, we asked the participants about their demographics, ai literacy levels and several other reasoning factors after the main set of questions, and these questions can be found in \S\ref{sssec:a-questions} The main questions for the second study can be found in Table \ref{app:part-2-questions}. The participant demographics for the second study can be found in Tables \ref{app:demographics-1-jobs-p2} to \ref{app:demographics-3-personal-p2}. The distribution of each use case within each scenario (Labor Replacement Use Cases and Personal Use Cases) for the second study is relatively well-balanced and can be found in Table \ref{tab:study2-use-case}.

\begin{table}[!hbpt]
    \centering
    \small
    \begin{tabularx}{\linewidth}{l|X}
    \toprule
       Use Case & Participants Allocated \\
    \midrule
       Personal Use Cases & \\
    \midrule
       Digital Medical Advice & 20 \\
       Customized Lifestyle Coach & 20 \\
       Personal Health Research & 19 \\
       Nutrition Optimizer & 21 \\
       Flavorful Swaps & 17 \\
    \midrule
       Labor Replacement Use Cases & \\
    \midrule
       Lawyer & 20 \\
       Elementary School Teacher & 22 \\
       IT Support Specialist & 17 \\
       Government Eligibility Interviewer & 19 \\
       Telemarketer & 23 \\
    \bottomrule
    \end{tabularx}
    \caption{Use Case allocation for Study 2. Specific participant numbers are listed for each use case.}
    \label{tab:study2-use-case}
\end{table}
\begin{table*}[!hbpt]
    \footnotesize
    \centering
    \begin{tabularx}{\linewidth}{l|X|l}
    \toprule
         Question ID & Question & Answer Type \\
         \midrule
         \multicolumn{3}{l}{AI Perception Question (Before)}\\
         \midrule
         AI Perception Before & Overall, how does the growing presence of artificial intelligence (AI) in daily life and society make you feel? & 5 Point Likert Scale\\
         \midrule
         \multicolumn{3}{l}{Initial Decision/Usage}\\
         \midrule
         Q1 - Initial & Do you think a technology like this should be developed? & Yes/No\\
         Q2 - Initial & How confident are you in your above answer? & 5 Point Likert Scale \\
         Q3 - Initial & If [Use Case] exists, would you ever use its services (answer yes, even if you think you would use it very infrequently)? & Yes/No \\
         Q4 - Initial & How confident are you in your above answer? & 5 Point Likert Scale \\
         \midrule
         \multicolumn{3}{l}{Benefits of Developing Use Case}\\
         \midrule
         Q1 - BDev & How will [Use Case] positively impact individuals? & Text \\
         Q2 - BDev & Which groups of people do you think would benefit the most from the above positive impacts? (You can list more than one group.) & Text \\
         Q3 - BDev & How beneficial would [Use Case] be if it had the above positive impacts? & 9 Point Likert Scale \\
         \midrule
         \multicolumn{3}{l}{Malicious Uses of Developing Use Case}\\
         \midrule
         Q1 - HDev & Please complete the following: [Use Case] could have a negative impact if it was used to... & Text \\
         Q1 - HDev & What would be the negative impact of the above malicious or unintended uses? & Text \\
         Q2 - HDev & Which groups of people do you think would be harmed the most by the above malicious or unintended uses? (You can list more than one group.) & Text \\
         Q3 - HDev & How harmful would [Use Case] be if it had the above negative impacts? & 9 Point Likert Scale \\
         \midrule
         \multicolumn{3}{l}{Failures of Developing Use Case}\\
         \midrule
         Q1 - HDevF & Please complete the following: If [Use Case] failed to do its intended task properly, fully, and accurately, it could have a negative impact if it... & Text \\
         Q1 - HDevF & What would be the negative impact of those failure cases? & Text \\
         Q2 - HDevF & Which groups of people do you think would be harmed the most by the above failure cases? (You can list more than one group.) & Text \\
         Q3 - HDevF & How harmful would [Use Case] be if it had the above negative impacts? & 9 Point Likert Scale \\
         \midrule
         \multicolumn{3}{l}{Benefits of Not Developing Use Case}\\
         \midrule
         Q1 - BNonDev & Please complete the following: Not having [Use Case] would be beneficial because... & Text \\
         Q2 - BNonDev & Which groups of people do you think would benefit the most by banning or not developing [Use Case]?
(You can list more than one group.) & Text \\
         Q3 - BNonDev & How beneficial would it be if [Use Case] was banned or not developed and it had the above positive impact? & 9 Point Likert Scale \\
         \midrule
         \multicolumn{3}{l}{Harms of Not Developing Use Case}\\
         \midrule
         Q1 - HNonDev & Please complete the following: Not having [Use Case] would be harmful because... & Text \\
         Q2 - HNonDev & Which groups of people do you think would be harmed the most by banning or not developing [Use Case]? (You can list more than one group.) & Text \\
         Q3 - HNonDev & How harmful would it be if [Use Case] was banned or not developed and it had the above negative impacts? & 9 Point Likert Scale \\
         \midrule
         \multicolumn{3}{l}{Final Decision/Usage}\\
         \midrule
         Q1 - Final & Do you think a technology like this should be developed? & Yes/No\\
         Q2 - Final & How confident are you in your above answer? & 5 Point Likert Scale \\
         Q3 - Final - Y & Please elaborate on your answer to the previous question: Do you think a technology like this should be developed?: [Q1 - Final Answer] & Text \\
         Q3 - Final - N & Please elaborate on your answer to the previous question: Do you think a technology like this should be developed?: [Q1 - Final Answer] & Text \\
         Q4 - Final - Y & Under what circumstances would you switch your decision from [Q1 - Final Answer] should be developed to should not be developed? & Text \\
         Q4 - Final - N & Under what circumstances would you switch your decision from [Q1 - Final Answer] should not be developed to should be developed? & Text \\ 
         Q5 - Final & If [Use Case] exists, would you ever use its services (answer yes, even if you think you would use it very infrequently)? & Yes/No \\
         Q6 - Final & How confident are you in your above answer? & 5 Point Likert Scale \\
         \midrule
         \multicolumn{3}{l}{AI Perception Question (After)}\\
         \midrule
         AI Perception After & Before we continue, we’d like to get your thoughts on AI one more time. Overall, how does the growing presence of artificial intelligence (AI) in daily life and society make you feel? & 5 Point Likert Scale \\
    \bottomrule
    \end{tabularx}
    \caption{Study 2 Specific Question. The placeholder [Use Case] is used in place of the 10 use cases chosen for the studies.}
    \label{app:part-2-questions}
\end{table*}
\begin{table*}[!htpb]
    \footnotesize
    \centering
    \begin{tabular}{{ll|ll|ll|ll}}
    \toprule
         \textbf{Racial Identity} & \textbf{\textit(N) (\%)} & \textbf{Age} & \textbf{\textit{N} (\%)} & \textbf{Gender Identity} & \textbf{\textit{N} (\%)} & \textbf{Education} & \textbf{\textit{N} (\%)} \\
         \midrule
        White or Caucasian & 32 (31.4) & 45-54 & 32 (31.4) & Man & 49 (49.0) & Bachelor’s degree & 44 (43.1)\\
        Black or African American & 25 (24.5) & 25-34 & 29 (28.4) & Non-male & 51 (51.0) & Graduate degree$^{*}$ & 18 (17.6)\\
        Asian & 18 (17.6) & 35-44 & 12 (11.8) & & & Some college $^{*}$ & 18 (17.6)\\
        Mixed & 15 (14.7) & 55-64 & 12 (11.8) & & & High school diploma$^{*}$ & 15 (14.7)\\
        Other & 12 (11.8) & 18-24 & 9 (8.8) & & & Associates degree$^{*}$ & 7 (6.9)\\
         & 2 (0.7) & 65+ & 8 (7.8) & & & Some high school$^{*}$ & 0 (0.0)\\
    \bottomrule
    \end{tabular}
    \caption{Labor Replacement Study 2 Survey: Racial, age, gender identities and education level of participants. Asterisk (*) denotes labels shortened due to space.}
    \label{app:demographics-1-jobs-p2}
\end{table*}
\begin{table*}[htpb]
    \centering
    \footnotesize
    \begin{tabular}{ll|ll|ll|ll}
    \toprule
         \textbf{Minority/Disadvantaged Group} & \textbf{\textit(N) (\%)} & \textbf{Transgender} & \textbf{\textit{N} (\%)} & \textbf{Sexuality} & \textbf{\textit{N} (\%)} & \textbf{Political Leaning} & \textbf{\textit{N} (\%)} \\
         \midrule
No & 56 (54.9) & No & 97 (95.1) & Heterosexual & 76 (74.5) & Liberal & 37 (36.3)\\
Yes & 46 (45.1) & Yes & 4 (3.9) & Others & 26 (25.5) & Moderate & 27 (26.5)\\
 &  & Prefer not to say & 1 (1.0) & & & Conservative & 17 (16.7)\\
 &  &  &  &  &  & Strongly liberal & 16 (15.7)\\
 &  &  &  &  &  & Strongly conservative & 4 (3.9)\\
 &  &  &  &  &  & Prefer not to say & 1 (1.0)\\
\bottomrule
    \end{tabular}
    \caption{Labor Replacement Study 2 Survey: Additional demographic identities}
    \label{app:demographics-2-jobs-p2}
\end{table*}
\begin{table*}[htpb]
    \centering
    \footnotesize
    \begin{tabularx}{\textwidth}{Xl|Xl|Xl|Xl}
    \toprule
        \textbf{Longest Residence} & \textbf{\textit(N) (\%)} & \textbf{Employment} & \textbf{\textit{N} (\%)} & \textbf{Occupation (Top 10)} & \textbf{\textit{N} (\%)} & \textbf{Religion} & \textbf{\textit{N} (\%)} \\
    \midrule
United States of America & 96 (94.1) & Employed, 40+ & 44 (43.1) & Other & 34 (33.3) & Christian & 38 (37.3)\\
Others & 6 (5.9) & Employed, 1-39 & 28 (27.5) & Educational Services & 11 (10.8) & Agnostic & 19 (18.6)\\
 &  & Not employed, looking for work & 19 (18.6) & Health Care and Social Assistance & 10 (10.0) & Catholic & 18 (17.6)\\
 &  & Retired & 4 (3.9) & Information & 8 (7.8) & Nothing in particular & 12 (11.8)\\
 &  & Not employed, NOT looking for work & 3 (2.9) & Prefer not to answer & 8 (7.8) & Atheist & 7 (6.9)\\
 &  & Other: please specify & 3 (2.9) & Retail Trade & 7 (6.9) & Muslim & 3 (2.9)\\
 &  & Disabled, not able to work & 1 (1.0) & Finance and Insurance & 7 (6.9) & Something else, Specify & 3 (2.9)\\
 &  & Prefer not to disclose & 0 (0.0) & Professional, Scientific, and Technical Services & 6 (5.9) & Jewish & 1 (1.0)\\
 &  &  &  & Manufacturing & 6 (5.9) & Hindu & 1 (1.0)\\
 &  &  &  & Administrative and support and waste management services & 5 (4.9) & Buddhist & 0 (0.0)\\
 \bottomrule
    \end{tabularx}
    \caption{Labor Replacement Study 2 Survey: Additional demographic identities. The Occupation category was capped at the top 10 for brevity, with the remaining occupations merged together with the Other: please specify option.}
    \label{app:demographics-3-jobs-p2}
\end{table*}
\begin{table*}[!htpb]
    \footnotesize
    \centering
    \begin{tabular}{{ll|ll|ll|ll}}
    \toprule
         \textbf{Racial Identity} & \textbf{\textit(N) (\%)} & \textbf{Age} & \textbf{\textit{N} (\%)} & \textbf{Gender Identity} & \textbf{\textit{N} (\%)} & \textbf{Education} & \textbf{\textit{N} (\%)} \\
         \midrule
        White or Caucasian & 35 (36.1) & 45-54 & 29 (29.9) & Non-male & 50 (51.5) & Bachelor’s degree & 33 (34.0)\\
        Black or African American & 22 (22.7) & 25-34 & 22 (22.7) & Man & 47 (48.5) & Graduate degree$^{*}$ & 24 (24.7)\\
        Asian & 19 (19.6) & 55-64 & 17 (17.5) & & & Some college $^{*}$ & 18 (18.6)\\
        Mixed & 13 (13.4) & 35-44 & 17 (17.5) & & & High school diploma$^{*}$ & 12 (12.4)\\
        Other & 8 (8.2) & 18-24 & 6 (6.2) & & & Associates degree$^{*}$ & 10 (10.3)\\
         & 2 (0.7) & 65+ & 6 (6.2) & & & Some high school$^{*}$ & 0 (0.0)\\
         &  & Prefer not to disclose & 0 (0.0) & & & & \\
    \bottomrule
    \end{tabular}
    \caption{Personal Use Cases Study 2 Survey: Racial, age, gender identities and education level of participants. Asterisk (*) denotes labels shortened due to space.}
    \label{app:demographics-1-personal-p2}
\end{table*}
\begin{table*}[htpb]
    \centering
    \footnotesize
    \begin{tabular}{ll|ll|ll|ll}
    \toprule
         \textbf{Minority/Disadvantaged Group} & \textbf{\textit(N) (\%)} & \textbf{Transgender} & \textbf{\textit{N} (\%)} & \textbf{Sexuality} & \textbf{\textit{N} (\%)} & \textbf{Political Leaning} & \textbf{\textit{N} (\%)} \\
         \midrule
No & 50 (51.5) & No & 93 (95.9) & Heterosexual & 76 (78.4) & Liberal & 34 (35.1)\\
Yes & 47 (48.5) & Yes & 4 (4.1) & Others & 21 (21.6) & Moderate & 26 (26.8)\\
 &  & Prefer not to say & 0 (0.0) & & & Strongly liberal & 17 (17.5)\\
 &  &  &  &  &  & Conservative & 13 (13.4)\\
 &  &  &  &  &  & Strongly conservative & 6 (6.2)\\
 &  &  &  &  &  & Prefer not to say & 1 (1.0)\\
\bottomrule
    \end{tabular}
    \caption{Personal Use Cases Study 2 Survey: Additional demographic identities}
    \label{app:demographics-2-personal-p2}
\end{table*}
\begin{table*}[htpb]
    \centering
    \footnotesize
    \begin{tabularx}{\textwidth}{Xl|Xl|Xl|Xl}
    \toprule
        \textbf{Longest Residence} & \textbf{\textit(N) (\%)} & \textbf{Employment} & \textbf{\textit{N} (\%)} & \textbf{Occupation (Top 10)} & \textbf{\textit{N} (\%)} & \textbf{Religion} & \textbf{\textit{N} (\%)} \\
    \midrule
United States of America & 95 (97.9) & Employed, 40+ & 44 (45.4) & Other & 36 (37.1) & Christian & 43 (44.3)\\
Others & 2 (2.1) & Employed, 1-39 & 23 (23.7) & Health Care and Social Assistance & 13 (13.4) & Agnostic & 12 (12.4)\\
 &  & Not employed, looking for work & 9 (9.3) & Information & 9 (9.3) & Atheist & 12 (12.4)\\
 &  & Other: please specify & 7 (7.2) & Finance and Insurance & 8 (8.2) & Catholic & 10 (10.3)\\
 &  & Retired & 6 (6.2) & Prefer not to answer & 6 (6.2) & Nothing in particular & 8 (8.2)\\
 &  & Disabled, not able to work & 5 (5.2) & Retail Trade & 6 (6.2) & Muslim & 4 (4.1)\\
 &  & Not employed, NOT looking for work & 2 (2.1) & Manufacturing & 5 (5.1) & Something else, Specify & 4 (4.1)\\
 &  & Prefer not to disclose & 1 (1.0) & Educational Services & 5 (5.1) & Buddhist & 2 (2.1)\\
 &  &  &  & Arts, Entertainment, and Recreation & 5 (5.1) & Hindu & 1 (1.0)\\
 &  &  &  & Accommodation and Food Services & 4 (4.1) & Jewish & 1 (1.0)\\
 \bottomrule
    \end{tabularx}
    \caption{Personal Use Cases Study 1 Survey: Additional demographic identities. The Occupation category was capped at the top 10 for brevity, with the remaining occupations merged together with the Other: please specify option.}
    \label{app:demographics-3-personal-p2}
\end{table*}

\subsection{Results}\label{sssec:a-results}
To explore the possible impact of explicitly weighing harms and benefits of a use case on participant's decision, we analyzed the participant's judgment of acceptability before and after explicit weighing of harms and benefits (Study 2; see Table\ref{app:part-2-questions} for details on questions asked). The Type III ANOVA with Satterthwaite's method for measurement time (before, after) indicated a marginally significant effect \( F(1, 201.05) = 3.371, p = 0.0678 \) on usage judgment weighed by confidence, which suggests that explicit harms and benefits weighing may have an influence, albeit not at conventional significance levels. We further analyzed reasoning effect on each subset of data pertinent to each use case through a mixed effects regression model with judgment metric as a dependent variable and measurement time as an independent variable with random effect from subject. Interestingly, the result was significant for Customized Lifestyle Coach AI across different judgments including, existence ($\beta=-0.40,SE=0.18,p<.05$), confidence-weighed existence ($\beta=-1.05,SE=0.51,p<.05$), and confidence-weighed usage judgments ($\beta=-0.75,SE=0.38,p<.05$). Explicit weighing also had a significant effect on confidence of existence judgment for Digital Medical Advice AI ($\beta=0.30,SE=0.11,p<.01$). The negative coefficients for Customized Lifestyle AI suggests that weighing harms and benefits caused participants to lower acceptance and positive coefficient to confidence on judgments on Digital Medical AI suggests that weighing harms and benefits solidified decisions. These diverging effects signify an interesting interaction between use cases and explicit weighing of harms and benefits.



\end{document}